\documentclass[12pt]{article}
\usepackage{amsmath,graphicx}
\usepackage{bbm}
\usepackage{amssymb}
\usepackage{epsf}
\usepackage{epsfig}
\usepackage{psfrag}
\usepackage{rotating}
\usepackage{color}

\usepackage{html}

\newcommand{\barms}{$\overline{\mathrm{MS}}$}

\newcommand{\lsim}{
\mathrel{\hbox{\rlap{\hbox{\lower4pt\hbox{$\sim$}}}\hbox{$<$}}}}

\newcommand{\gsim}{
\mathrel{\hbox{\rlap{\hbox{\lower4pt\hbox{$\sim$}}}\hbox{$>$}}}}

\newcommand{\gev}{\, {\rm GeV}}

\newcommand{\tev}{\, {\rm TeV}}
\newcommand{\ev}{\, {\rm eV}}

\newcommand{\sle}{\stackrel{<}{{}_\sim} }

\newcommand{\bea}{\begin{eqnarray}}
\newcommand{\eea}{\end{eqnarray}}

\newcommand{\be}{\begin{equation}}
\newcommand{\ee}{\end{equation}}
\newcommand{\bi}{\begin{itemize}}
\newcommand{\ei}{\end{itemize}}
\newcommand{\ord}{{\cal O}}
\newcommand{\RE}{{\rm Re}}
\newcommand{\IM}{{\rm Im}}

\renewcommand{\baselinestretch}{1.2}

\begin{document}
\begin{titlepage}
\vspace*{-0.5truecm}

\begin{flushright}
TUM-HEP-645/06\\
PITHA-06/08
\end{flushright}

\vspace*{0.3truecm}

\begin{center}
\boldmath

{\Large{\bf Another Look at Minimal Lepton Flavour Violation,  
\vspace{0.3truecm}
 $l_i\to l_j\gamma$,  
Leptogenesis
and the Ratio $M_\nu/\Lambda_{\rm LFV}$}}
\unboldmath
\end{center}


\begin{center}
\small

{\bf Gustavo C. Branco,${}^{a,b}$ Andrzej J. Buras,${}^a$ 
Sebastian  J\"ager,${}^c$ 
\\ 
\vspace{0.3truecm}
Selma Uhlig${}^a$ and Andreas Weiler${}^a$}

\vspace{0.9truecm}

${}^a$ {\sl Physik Department, Technische Universit\"at M\"unchen,
D-85748 Garching, Germany}\\
${}^b$ {\sl Centro de F\'{\i}sica Te\'{o}rica de Part\'{\i}culas, 
Departamento de F\'{\i}sica,
Instituto Superior T\'{e}cnico, Av. Rovisco
Pais, 1049-001 Lisboa, Portugal}\\
${}^c${\sl Institut  f{\"u}r Theoretische Physik E, RWTH Aachen, D-52056
Aachen, Germany}

\normalsize

\end{center}
\vspace{0.3truecm}
\renewcommand{\baselinestretch}{1}
\begin{abstract}

 We analyze lepton flavour violation (LFV), as well as generation of the observed
baryon-antibaryon asymmetry of the Universe (BAU) within a generalized 
minimal lepton flavour violation (MLFV) framework where we allow for CP 
violation both at low and high energies. The generation of BAU is obtained 
through  {\it radiative resonant leptogenesis} (RRL), where starting with 
three 
exactly
degenerate right--handed neutrinos at $\Lambda_{\rm GUT}$, we demonstrate
explicitly within the SM and the MSSM that 
the splittings between their masses at 
the see-saw scale $M_\nu$, generated by
renormalization group effects, are sufficient for a successful leptogenesis
for $M_\nu$ even as low as $10^6\gev$.
The inclusion of flavour effects plays an important role in this
result
and can lead to the observed BAU even in the absence
of CP violation beyond the PMNS phases.
The absence of a stringent lower bound on $M_\nu$ in this type of leptogenesis
allows to easily satisfy present and  near future  
upper bounds on 
$\mu\to e\gamma$ and other charged lepton flavour violating (LFV) processes
even for $\Lambda_{\rm LFV}=\ord(1\tev)$.
We find, that the 
MLFV framework in the presence of heavy right-handed neutrinos and 
leptogenesis is  not as predictive 
as MFV in the quark sector and  
point out that without a specific MLFV model, there is a 
rich spectrum of possibilities for charged LFV processes and 
for their correlation 
with low energy neutrino physics and the LHC physics, 
even if the constraint from the observed BAU is taken into 
account.  While certain qualitative features of our analysis confirm findings of Cirigliano et al., at the quantitative level we find phenomenologically important differences. We explain the origin of these differences.

\renewcommand{\baselinestretch}{1.2}
\end{abstract}

\thispagestyle{empty}\vbox{}

\end{titlepage}

\thispagestyle{empty}
\vbox{}

\setcounter{page}{1}
\pagenumbering{arabic}

\section{Introduction}
One of the attractive and predictive frameworks for the description of
flavour changing processes in the quark sector is the so--called Minimal
Flavour Violation (MFV) hypothesis \cite{UUT,AMGIISST} in which the 
Standard Model (SM) quark 
Yukawa couplings are
the only sources of flavour changing and in particular CP-violating
processes.\footnote{For earlier discussions of this hypothesis see \cite{Chivukula:1987py}.}

If only one Higgs doublet is involved in the spontaneous breaking
of the underlying gauge symmetry, all flavour changing charged and neutral
current processes are governed in the MFV framework
 by the CKM matrix \cite{CKM} and the 
relevant local
operators are only those present in the SM. As demonstrated in 
\cite{MFVB}, the
existing data on $B^0_{d,s}-\bar B^0_{d,s}$ mixing, $\varepsilon_K$, 
$B\to X_s\gamma$, $B\to X_s l^+l^-$ and $K^+\to\pi^+\nu\bar\nu$ and the value
of the angle $\beta$ in the unitarity triangle from the mixing induced CP
asymmetry in  $B\to\psi K_S$ imply within this framework very stringent
bounds on all rare $K$ and $B$ decay branching ratios. Consequently, 
substantial
departures from the SM predictions are not expected if MFV with one Higgs 
doublet is the whole
story. 

If two Higgs doublets, like in the MSSM, are involved and the ratio of
the corresponding vacuum expectation values $v_2/v_1\equiv \tan\beta$ is
large, significant departures from the SM predictions for certain decays are
still possible within the MFV framework \cite{AMGIISST} in spite of the 
processes being
governed solely by the CKM matrix. The most prominent examples are the decays
$B_{d,s}\to\mu^+\mu^-$ with a subset of references given in \cite{LTANB}. 
The prime reason for these novel 
effects is the
appearance of new scalar operators that are usually strongly suppressed
within the SM and MFV models at low $\tan\beta$, but can become important and
even dominant for large $\tan\beta$. The improved data on
$B_{d,s}\to\mu^+\mu^-$, expected to come in this decade from Tevatron and 
LHC,
will tell us whether MFV models with large $\tan\beta$ are viable.

One of the important virtues of the MFV in the quark sector are the relations
\cite{UUT,BBGT} between the ratios of various branching ratios and the CKM
parameters measured in low energy processes that have universal character 
and are independent of the details of the specific MFV model. An example is 
the universal unitarity triangle common to all MFV models \cite{UUT}. 
But also the fact 
that each branching ratio can be expressed in terms of the CKM parameters 
and quark masses measured at the electroweak scale or lower energy scales 
makes this scenario to be a very predictive framework. Moreover, neither fine 
tuning nor the introduction of unnaturally high scales of new physics are
required to make this scenario consistent with the available data.

The MFV scenario in the quark sector in question, although simple and
elegant, suffers from the following problem. In the absence of new complex
phases beyond the CKM phase, it cannot accommodate the observed size of the 
baryon asymmetry of the
universe (BAU) to be denoted by $\eta_B$ in what follows. 
The question then arises, whether one could still explain the
right size of $\eta_B$ within the MFV context by considering simultaneously the
lepton sector, where the BAU can in principle be explained with the help of 
leptogenesis \cite{Fukugita:1986hr,Buchmuller:2005eh}. While this is the 
most natural possibility, other directions could be explored in principle.

Before addressing this question let us summarize what is known in the
literature about the MFV in the lepton sector. Last year, Cirigliano,
Grinstein, Isidori and Wise 
\cite{MLFV} in an interesting paper formulated MFV in the lepton sector
(MLFV) both with
the minimal field content and with the extended field content, where three 
degenerate right-handed heavy neutrinos $\nu^i_R$ with masses $M_\nu^i$ are 
added to the SM fields
and the see-saw mechanism \cite{Minkowski:1977sc} is responsible for the
generation of the light neutrino masses with the see-saw scale denoted 
by $M_\nu$ in what follows.
Analyzing charged lepton flavour violating (LFV) processes, 
like $\mu\to e\gamma$ and
$\mu\to e$ conversion in nuclei in these two scenarios in the absence of 
CP violation, they reached two interesting
conclusions:
\begin{itemize}
\item
 Measurable rates for LFV processes within MLFV are only obtained when 
the scale
for total lepton number violation $(\Lambda_{\rm LN}=\ord(M_\nu))$ is by many 
orders of
magnitude, typically a factor $10^7-10^9$, larger than the scale of charged 
lepton 
flavour violation $(\Lambda_{\rm LFV})$. 
\item
Similarly to MFV in
the quark sector, the ratios of various LFV rates like 
$B(\mu\to e\gamma)/B(\tau\to \mu\gamma)$ are unambiguously determined in
terms of neutrino masses and mixing angles measured in low energy processes.
\end{itemize}
Various phenomenological aspects of 
MLFV, as formulated in \cite{MLFV}, have been subsequently discussed in 
\cite{MLFV2}.

 The MLFV framework  in \cite{MLFV} does not include 
CP violation, neither at low energy nor at high energy,
being a necessary ingredient in the generation of the BAU.
Moreover, possible renormalization group effects between the low energy 
scale $\ord(M_Z)$ and the high energy scales, like the see-saw scale 
$M_\nu$ and the GUT scale $\Lambda_{\rm GUT}$, have not been taken into
account in \cite{MLFV}. It is then natural to ask: 
\begin{itemize}
\item
whether a successful 
leptogenesis is at all possible within a MLFV framework in which flavour 
violation is governed solely by Yukawa couplings,
\item
how the findings of 
\cite{MLFV} are modified, when CP violation at low and high energy and 
the renormalization group effects (RGE) in question are taken into account,
\item
whether a successful leptogenesis in the MLFV framework puts stringent 
constraints on charged LFV processes.
\end{itemize}

The main goal of our paper is to answer these three questions. In fact, 
as we will demonstrate explicitely in Section 5, it is possible to obtain 
the correct size of $\eta_B$ in the MLFV framework with three  
heavy right-handed neutrinos that are assumed to be 
degenerate in mass at $\Lambda_{\rm GUT}$. Other choices for this scale could be
considered but $\Lambda_{\rm GUT}$ seems to be the most natural one. 
The breakdown of this degeneracy through RGE, 
that are governed by Yukawa couplings, combined with new sources of 
CP-violation in the heavy neutrino sector allows to obtain the correct 
size of $\eta_B$ in the framework of the resonant leptogenesis in
particular when flavour effects are taken into account. As this 
type of leptogenesis is generated  here radiatively and not put by hand 
as done in most literature sofar, we will call this scheme {\it radiative
resonant leptogenesis} (RRL) in what follows. 

The fact that within the MLFV framework one is naturally led to RRL, has 
significant implications on charged LFV processes, which could in principle be
used to distinguish this scenario from other extensions of the SM.
In particular, while other types of leptogenesis, with hierarchical 
right-handed heavy neutrinos, imply generally rather stringent lower 
bounds for the lightest $\nu_R^i$ mass, in the ballpark of $\ord(10^{8}\gev)$ 
or higher, the values of $M_\nu$ in RRL are allowed to be by many orders 
of magnitude lower. As the branching ratios for $l_i\to l_j\gamma$ are
proportional to $M_\nu^2/\Lambda_{\rm LFV}^4$ \cite{MLFV}, it is relatively 
easy to satisfy the present and in the near future available upper 
bounds on these processes by simply choosing sufficiently small value 
of $M_\nu$. Conversely, by choosing $M_\nu$ to be larger than say 
$10^{12}\gev$,
it is in principle possible to obtain the values of $B(\mu\to e\gamma)$ 
close to expected bounds from PSI even if $\Lambda_{\rm LFV}$ is as high as 
$100\tev$. This means that non-observation of $\mu\to e\gamma$ with the 
rate $10^{-13}$ at PSI will not necessarily imply within the general 
MLFV framework that $\Lambda_{\rm LFV}$ is very high. Conversely, the observation 
of $\mu\to e\gamma$ will not necessarily imply LFV physics at scales 
$\ord(1\tev)$. In other words, without a specific MLFV model there is a 
rich spectrum of possibilities for charged LFV processes within the general 
MLFV framework, even if the constraint from $\eta_B$ is taken into account.

Thus one of the main messages of our paper is the realization that the 
MLFV framework in the presence of heavy right-handed neutrinos and 
leptogenesis is clearly  not as  
predictive as MFV in the quark sector. This is also related 
to the fact that new physics, even lepton conserving one, that could 
be present between energy scales $M_Z$ and $M_\nu$, could have in principle 
an important impact on various observables, like $B(l_i\to l_j\gamma)$, 
 through RGE.

In the advanced stages of our project a paper by Cirigliano, Isidori and Porretti
\cite{CIP06} appeared, in which the idea of the incorporation of leptogenesis 
into the MLFV framework has been put forward in the literature for the first time and its implications for 
charged LFV 
processes have been analyzed in detail. 
While in agreement with the general
predictions of \cite{CIP06} we find that successful leptogenesis is possible
for high scales $M_\nu \ge 10^{12}$ GeV, and contrary to that paper our detailed
numerical analysis demonstrates that this is also true for much lower scales,
weakening the implications for charged LFV processes found in that paper.
Most importantly 
we do not confirm the lower bound of $10^{12}\gev$ for $M_\nu$ found by these
authors which has significant implications for charged LFV processes 
as stressed
above. 
The inclusion 
of flavour effects in the leptogenesis in our paper, that has been 
left out in \cite{CIP06} and the use of approximate formulae in that
paper as opposed to a full numerical analysis present here, brings in
significant differences in these two analyses for $M_\nu \le 10^{12}$
GeV. We will summarize the agreements and differences
between \cite{CIP06} and us in Section~\ref{sec:comparison}.

At this stage it is worth also mentioning that there may be other 
equally reasonable definitions of MLFV. In this paper, we will only 
consider a conservative generalization of the initial proposal for MLFV~\cite{MLFV}.  
However, it is clear that one may have other well motivated but
different proposals for MLFV. In particular one should keep in mind that
within the seesaw mechanism neutrinos acquire a mass in a manner which
differs significantly from the one in the quark sector. In fact it has been 
suggested \cite{Altarelli:1999dg} that the fact that neutrino masses arise from the seesaw 
mechanism is the key point in understanding why leptonic mixing is large, 
in contrast with small quark mixing. Therefore a reasonable definition of 
MLFV may differ from MFV in the quark sector. In our opinion, only in the 
presence of a theory of lepton flavour, where Yukawa couplings would be 
constrained by family symmetries, can one define in a unique way what 
MLFV should be. The question of different definitions of MLFV has been 
recently addressed in an interesting paper by Davidson and Palorini~\cite{Davidson:2006bd}.

Our analysis involves several points and it is useful to list them one by one
already at this stage.

\begin{itemize}
\item
As already stated above, in the
framework of the MLFV the right-handed heavy neutrinos are assumed to be 
degenerate in mass at some high energy scale 
 in order to exclude possible new sources of
flavour violation. However, the exact degeneracy of $M^i_\nu$ is not RG 
invariant and can only be true at a single scale which we choose to be 
$\Lambda_{\rm GUT}=\ord(10^{16}\gev)$. RGE between 
$\Lambda_{\rm GUT}$ and the see-saw scale $M_\nu\ll \Lambda_{\rm GUT}$ break
the degeneracy between $M^i_\nu$ at $M_\nu$, a welcome result 
for leptogenesis that vanishes in the limit of degenerate $M^i_\nu$.
This structure is the basis of the so called {\it radiative leptogenesis}
\cite{GonzalezFelipe:2003fi,Turzynski:2004xy}
that has been first  considered in the case of two degenerate neutrinos 
in \cite{GonzalezFelipe:2003fi,Turzynski:2004xy,Branco05}.
An important ingredient of this framework is the {\it resonant 
leptogenesis} \cite{Pilaftsis:2005rv,Pilaftsis:1997jf,Pilaftsis:2003gt}. Therefore we will call 
this framework RRL as stated above. 
Our analysis is one of the first that considers the case
of three degenerate neutrinos and includes flavour effects in RRL. 

\item
As the values of the light neutrino masses and of the parameters of the PMNS
mixing matrix \cite{PMNS}, that enter the formulae for charged LFV processes,
are not to be evaluated at the low energy scale but at the high energy scale
$M_\nu$, the MLFV relations between neutrino masses,
mixing angles and rates for charged LFV processes presented in \cite{MLFV} 
can be 
in principle significantly modified through the RGE
between the $M_Z$ and $M_\nu$ scales, changing the conclusions
about the value of the ratio $M_\nu/\Lambda_{\rm LFV}$ necessary
to obtain visible charged LFV rates. 
While it is conceivable that in certain MLFV
scenarios RGE could be neglected, the example of the MSSM with a large
$\tan\beta$, presented in this context in \cite{Petcov:2005yh}, 
shows that the RGE in question could in principle modify 
$B(l_i\to l_j\gamma)$ by a few orders of magnitude.
\item
The requirement of a successful BAU with the help of 
leptogenesis and in fact in general, necessarily brings into play CP 
violation. 
Neglecting flavour effects in the Boltzmann equations,
the relevant CP violation is encoded in a complex orthogonal matrix $R$ in
the parameterization of $Y_\nu$ by Casas and Ibarra \cite{CAIB}.

As analyzed already in several papers in the context
of supersymmetric models, the size of 
the imaginary parts of $R$, crucial for generating
the observed BAU in the framework of
leptogenesis, can change the rates of charged LFV processes by several 
orders of
magnitude. See, in particular \cite{Pascoli:2003rq}, but also 
\cite{
Petcov:2005yh,CAIB,Kanemura:2005cq,Kanemura:2005it,Ellis:2002eh,Petcov:2005jh,Deppisch:2005rv}. We note that when flavour effects are important, the generation of the BAU could be possible without complex phases in $R$~\cite{Abada:2006fw,Nardi:2006fx}.
\item
The inclusion of CP violation at low energy with the help of the
non-vanishing phase $\delta_{\rm PMNS}$ \cite{PMNS} has
 only a moderate impact 
on the results in \cite{MLFV} but in the presence of a complex matrix 
$R$ (see above) 
 non-vanishing Majorana phases in the PMNS matrix 
can modify the results for LFV processes in \cite{MLFV}  
both directly and indirectly through 
RGE mentioned above. The numerical studies in 
\cite{Kanemura:2005cq,Kanemura:2005it,Petcov:2005yh}   
show that such effects can be in principle significant.
\end{itemize}

Our paper is organized as follows. In Section~\ref{sec:framework} we present the generalization
of the formulation of MLFV given in \cite{MLFV} by 
 including low energy CP violation in the leptonic sector with the help of 
the
PMNS matrix \cite{PMNS} 
and the high energy CP violation necessary for
 the leptogenesis of BAU. 
The parametrization of the neutrino Yukawa coupling $Y_\nu$ of Casas and
Ibarra \cite{CAIB}
turns out to be  useful here.

In Section~\ref{sec:RGE} we analyze the issue of the breakdown of the mass degeneracy of 
heavy neutrinos by radiative corrections. 
This breakdown is necessary for leptogenesis to work
even if $R$ is complex. 
 Assuming then this scale to be
the grand unification (GUT) scale, we discuss the renormalization 
group
equations in the SM and the MSSM used to generate the splitting of $M_i$ at scales 
$\ord(M_\nu)$, where the heavy
neutrinos are integrated out. 
The results of this section are a very 
important ingredient of the leptogenesis that we consider in Section~\ref{sec:numerics} and in particular in~\ref{sec:leptogenesis}.
 
In Section~\ref{sec:numerics} as a preparation for Section~\ref{sec:leptogenesis}, we present some numerical aspects of the flavour changing radiative 
charged lepton decays 
$l_i\to l_j\gamma$ and of the CP asymmetries in the right-handed neutrino decays.

In Section~\ref{sec:leptogenesis}, the most important section of our paper, 
we  describe the scenario of radiative resonant leptogenesis in the case
of three quasi-degenerate right-handed Majorana neutrinos. 
In this context we include in our analysis recently discussed flavour 
effects that definitely cannot be neglected. The main result of this 
paper is the demonstration that the right value of $\eta_B$ can be 
obtained in this framework. A plot of $\eta_B$ versus $M_\nu$ demonstrates 
very clearly that already for $M_\nu$ as low as $10^6$ GeV leptogenesis becomes
effective and that flavour effects are important. We compare our
results with existing literature and explain why in contrast to
\cite{CIP06} we do not find a stringent lower bound on $M_\nu$.

Finally, we return to the $l_i\to l_j\gamma$ decays and use the 
knowledge collected in Sections~\ref{sec:RGE} and~\ref{sec:leptogenesis} to present a brief numerical
analysis of $\mu\to e \gamma$ that illustrates the points
made above.
We restrict our analysis to $\tan\beta\le 10$ so that RGE between $M_Z$ 
and $M_\nu$ are small and other effects can be transparently seen.

In Section~\ref{sec:comparison} we compare our analysis and our results with \cite{CIP06}.
We conclude in Section~\ref{sec:conclusions}.

\section{Basic Framework}
\label{sec:framework}
\setcounter{equation}{0}
\subsection{Preliminaries}
The discovery of neutrino oscillations provides evidence
for non-vanishing neutrino masses and leptonic mixing, leading to
lepton-flavour violation. In the SM, neutrinos are strictly 
massless, since Dirac masses cannot be constructed due to the absence of 
right-handed neutrinos, and left-handed Majorana masses are not generated 
due to exact $(B-L)$ conservation.

The simplest extension of the SM which allows for non-vanishing but naturally
small neutrino masses, consists of the addition of right-handed neutrinos
to the spectrum of the SM. This extension has the nice feature of 
establishing on the one hand a lepton quark symmetry and on the other hand
being naturally embedded in a grand unified theory like $SO(10)$. Since 
right-handed neutrinos are singlets under $ U(1)\times SU(2) \times SU(3)$,
Majorana neutrino masses $M_R$ should be included, with a mass scale $M_\nu$
which can be much larger than the scale $v$ of the electroweak symmetry 
breaking. Apart from $M_R$, Dirac neutrino mass terms $m_D$ are generated 
through 
leptonic Yukawa couplings upon gauge symmetry breaking. The presence of these 
two neutrino mass terms leads, through the seesaw mechanism 
\cite{Minkowski:1977sc}, to three light
neutrinos with masses of order $v^2/M_\nu$ and three heavy neutrinos with mass of
order $M_\nu$.
The decay of these heavy neutrinos can play a crucial role in the creation 
of a baryon asymmetry of the universe (BAU) through the elegant mechanism of
baryogenesis through leptogenesis
\cite{Fukugita:1986hr,Buchmuller:2005eh}. 
 In the presence of neutrino masses and 
mixing, one has, in general, both CP violation at low energies which can be 
detected through neutrino oscillations and CP violation at high energies
which is an essential ingredient of leptogenesis. The connection
between these two manifestations of CP violation can be established in the
framework of specific lepton flavour models.

In this paper, we study lepton-flavour violation in this extension of the SM,
assuming minimal lepton flavour violation (MLFV) but allowing for CP violation
both at low and high energies. The case of no leptonic CP violation
either at low or high energies, was considered in \cite{MLFV} where the 
suggestion of MLFV was first presented. The first discussion of CP violation at low and high energy in a MLFV 
 framework has been presented recently in \cite{CIP06}. We will compare the 
 results of this paper with ours in Section \ref{sec:comparison}

\subsection{Yukawa Couplings and Majorana Mass Terms}
We add then three right-handed neutrinos to the spectrum of the SM and consider
the following leptonic Yukawa couplings and right-handed Majorana mass terms:
\be
\mathcal{L}_Y=-\bar e_R Y_E \phi^{\dagger} L_L   - 
\bar \nu_R Y_{\nu} \tilde\phi L_L  + h.c.
\ee
\be
\mathcal{L}_M=-\frac{1}{2} \bar \nu ^c_R M_R \nu_R + h.c. ~,
\ee
where $Y_E$, $Y_{\nu}$ and $M_R$ are $3 \times 3 $ matrices in the lepton 
flavour space.
 In the limit $\mathcal{L}_Y=\mathcal{L}_M=0$ the Lagrangian of this minimal 
extension of 
the SM has a large flavour symmetry 
\be              \label{eq:flavourgroup}
 SU(3)_L \times SU(3)_E \times SU(3)_{\nu_R} \times U(1)_L
\times U(1)_E \times U(1)_{\nu_R},
\ee
 which reflects the fact that gauge interactions treat all 
flavours on equal footing. This large global symmetry is broken by the Yukawa
couplings $Y_E$, 
$Y_{\nu}$ and by the Majorana mass terms $M_R$. A transformation of the 
lepton fields:
\be              \label{eq:trafofields}
L_L \rightarrow V_L L_L, \quad e_R \rightarrow V_E e_R, \quad \nu_R 
\rightarrow V_{\nu_R} \nu_R
\ee
leaves the full Lagrangian invariant, provided the Yukawa couplings and the 
Majorana mass terms 
transform as:
\begin{eqnarray} 
Y_{\nu} &\rightarrow& Y_{\nu}^{'} =  V_{\nu_R} Y_{\nu} V_L^{\dagger},
\label{trafo1} \\
Y_E & \rightarrow &  Y_{E}^{'} =   V_{E} Y_{E} V_L^{\dagger},
\label{trafo2}\\
M_R &\rightarrow& M_R^{'} = V_{\nu_R}^* M_R V_{\nu_R}^T,
\label{trafo3}
\end{eqnarray}
which means that there is a large equivalent class of Yukawa coupling matrices 
and Majorana mass 
terms, related through (\ref{trafo1})-(\ref{trafo3}), 
which have the same physical
content. The MLFV 
proposal \cite{MLFV} consists of the assumption that the physics which 
generates 
lepton number violation,
leading to $M_R$, is lepton flavour blind, thus leading to an exactly 
degenerate eigenvalue 
spectrum for $M_R$, at a high-energy scale. As a result, in the MLFV 
framework, the Majorana mass 
terms break $SU(3)_{\nu_R}$ into $O(3)_{\nu_R}$.

\subsection{Leptonic Masses, Mixing and CP Violations}
Without loss of generality, one can choose a basis for the leptonic fields, 
where $Y_E$ and $M_R$
are diagonal and real. In this basis, the neutrino Dirac mass matrix 
$m_D = v Y_{\nu}$ is an
arbitrary complex matrix, therefore with nine moduli and nine phases. 
Three of these phases can
be eliminated by a rephasing of $L_L$. One is then left with six CP violating
phases. There are 
various classes of phenomena which depend on different combinations $m_D$, 
$m_D^{T}$,
$m_D^{\dagger}$ or equivalently $Y_\nu$, $Y^{ T}_\nu$ and $Y^\dagger_\nu$:
\begin{itemize}
\item [\bf A)] {\bf Leptonic mixing and CP violation at low energies:} 
Since we are working in the 
      basis where the charged lepton mass matrix is diagonal and real, 
leptonic mixing and CP 
      violation at low energies are controlled by the PMNS matrix $U_{\nu}$ 
\cite{PMNS}, 
which diagonalizes the 
      effective low energy neutrino mass matrix:
\be
U_{\nu}^{T}(m_{\nu})_{eff} U_{\nu} = d_{\nu},
\ee      
      where 
$d_{\nu} \equiv \mathrm{diag} (m_{1}, m_{2}, m_{3})$, with $m_i$ being
the masses of the light neutrinos and
\cite{Minkowski:1977sc}
\be                    \label{eq:mnueff}
 (m_{\nu})_{eff} = -v^2 Y_{\nu}^T D_R^{-1}Y_{\nu}, 
\ee
where $D_R$ denotes the diagonal matrix $M_R$ and $v=174$ GeV.
      In the case of MLFV, $D_R = M_\nu \mathbbm{1}$ and one obtains
at $M_\nu\approx \Lambda_{\rm LN}$
\be
(m_{\nu})_{eff} =- \frac{v^2}{M_\nu} Y_{\nu}^T Y_{\nu}.
\ee
Consequently $Y_{\nu}^T Y_{\nu}$ is the quantity that matters here.

\item[\bf B)] {\bf Lepton flavour violation:} 
The charged LFV depends on the other hand on the 
combination $ Y_{\nu}^{\dagger} Y_{\nu}$ with $Y_\nu$ again normalized at the
high energy scale $M_\nu$. We will return to this point in Section~\ref{sec:numerics}.

\item[\bf C)] {\bf CP violation relevant for leptogenesis:}
The generation of BAU through leptogenesis starts by the production of
a lepton asymmetry which is proportional to the CP asymmetry in the
decays of heavy Majorana neutrinos. This CP asymmetry involves the
interference between the tree-level amplitude and the one-loop vertex
and self-energy contributions. It has been shown \cite{Covi:1996wh} that the
CP asymmetry depends on the neutrino Yukawa couplings through the
combination $Y_{\nu}Y_{\nu}^{\dagger}$. Again as in classes A and B, 
$Y_\nu$ is evaluated here at the scale $M_\nu$. When flavour effects in the Boltzmann equations become important,
the non-summed products $(Y_{\nu})_{ik} (Y_{\nu})_{jk}^*$ corresponding to different
lepton flavours $k$ can attain relevance.

\end{itemize}

\subsection{An Useful Parametrization}
In order to analyze in a systematic way the above phenomena and study
the implied relations among low-energy lepton mixing data,
lepton flavour violation and leptogenesis in different scenarios classified
below, it is convenient to choose
an appropriate parametrization for $Y_{\nu}$. We use the following
parametrization \cite{CAIB} of the neutrino Yukawa couplings:
\begin{equation}\label{yukawaparamet}
(\sqrt{D_R})^{-1}\, Y_{\nu}=\frac{i}{v}\, R \,\sqrt{d_{\nu}}
\,U_{\nu}^{\dagger}\, ,
\end{equation}
where $R$ is an orthogonal complex matrix
($R^TR=R\,R^T=\mathbbm{1}$), 
$d_{\nu}=\text{diag}(m_{1},m_{2},m_{3})$ and
$D_R=\text{diag}(M_1,M_2,M_3)$.

It is instructive to count next the number of independent parameters
on both sides of  (\ref{yukawaparamet}). The left-hand side of
(\ref{yukawaparamet}) is an arbitrary $3\times 3$ complex matrix with
nine real parameters and six phases, since three of the
initial nine phases can be removed by rephasing $L_L$. It is clear
that the right-hand side of (\ref{yukawaparamet}) also has nine real
parameters and six phases. Indeed, $R$, $d_{\nu}$ and $U_{\nu}$ have each
three real parameters and moreover $R$ and $U_{\nu}$ have in addition
each three phases. We consider now the case where the right-handed
neutrinos are exactly degenerate, i.e. $D_R=M_{\nu}\mathbbm{1}$. We will show
that three of the real parameters of $R$ can be rotated away. Note that
any complex orthogonal matrix can be parametrized as
\begin{equation}
R=e^{A_1}e^{iA_2},
\end{equation}
with $A_{1,2}$ real and skew symmetric. Now in the degenerate case an
orthogonal rotation of $\nu_R \rightarrow O_R \nu_R$ leaves the
Majorana mass proportional to the unit matrix and defines a physically
equivalent reparametrization of the fields $\nu_R$. Choosing
$O_R=e^{A_1}$ we see immediately that
\begin{equation}
Y_{\nu}\rightarrow O_{R}^{\dagger} Y_{\nu}=
\frac{\sqrt{M_\nu}}{v}\, e^{-A_1}\, R\,
\sqrt{d_{\nu}}\, U_{\nu}^{\dagger}=\frac{\sqrt{M_\nu}}{v}\, e^{iA_2}\,
\sqrt{d_{\nu}}\, U_{\nu}^{\dagger}\, ,
\end{equation}
which shows that the physically relevant parameterization is given by
$R_{deg}=e^{iA_2}$.

Using the parameterization in (\ref{yukawaparamet}) one finds that the
matrix $Y_{\nu}^{T}Y_{\nu}$ which controls
 low-energy CP-Violation and
mixing can be written as follows
\begin{equation}
Y^{T}_\nu Y_{\nu}
=-\frac{1}{v^2}\, (U^\dagger_\nu)^{T}\, \sqrt{d_{\nu}}\,
R^{\rm T}\, D_R\,
R \, \sqrt{d_{\nu}}\, U_{\nu}^{\dagger}=-
\frac{M_\nu}{v^2}(U^\dagger_\nu)^{T}d_\nu
U^\dagger_\nu,
\end{equation}
where in the last step we have set $D_R=M_{\nu}\mathbbm{1}$.

On the other hand, the matrix
$Y_{\nu}^{\dagger}Y_{\nu}$ which controls charged LFV,
can be written as follows (see also \cite{Petcov:2005jh})
\begin{equation}\label{YdaggerY}
Y_{\nu}^{\dagger}Y_{\nu}=\frac{1}{v^2}\, U_{\nu}\, \sqrt{d_{\nu}}\,
R^{\dagger}\, D_R\,
R \, \sqrt{d_{\nu}}\, U_{\nu}^{\dagger}=\frac{M_{\nu}}{v^2}\,
U_{\nu}\, \sqrt{d_{\nu}}\, R^{\dagger}
R\, \sqrt{d_{\nu}}\,U_{\nu}^{\dagger}\, .
\end{equation}

Finally, the matrix $Y_{\nu}Y_{\nu}^{\dagger}$ which enters
in leptogenesis when flavor effects are not relevant is given by (see also \cite{Petcov:2005jh}):
\begin{equation}\label{YYdagger}
Y_{\nu}Y_{\nu}^{\dagger}=\frac{1}{v^2}\, \sqrt{D_R}\, R\, d_{\nu}\,
R^{\dagger}\, \sqrt{D_R}=\frac{M_{\nu}}{v^2}\, R\, d_{\nu}\, R^{\dagger}.
\end{equation}

We note that $Y^{T}_\nu Y_{\nu}$ depends only on $U_\nu$ and
$d_\nu$, while $ Y_{\nu} Y_{\nu}^\dagger$ relevant for the
leptogenesis only on $d_\nu$ and $R$. This means that CP violation at
low energy originating in the complex $U_\nu$ and the CP violation
relevant for leptogenesis are then decoupled
from each other and only the
mass spectrum of light neutrinos summarized by $d_\nu$ enters both
phenomena in a universal way.

In this respect the charged LFV, represented by 
(\ref{YdaggerY}), appears also interesting
as it depends on $d_\nu$, $U_\nu$ and $R$ and consequently can also 
provide
an indirect link between low energy and high energy CP violations and
generally a link between low and high energy phenomena.

\subsection{Classification}\label{sec:classification}
Having the parametrization of $Y_\nu$ in (\ref{yukawaparamet}) at hand we 
can now spell the difference between the analysis of \cite{MLFV} and ours 
in explicit terms. 
Indeed,
from the above considerations, it follows that possible relations
among phenomena A,B,C, discussed in Section 2.3, crucially depend on the 
assumptions one makes
about leptonic CP violation at low energies, as well as at high
energies. One may consider then separately the following four scenarios:
\begin{itemize}
\item {\bf Case 1:} No leptonic CP violation either at low or high
  energies. The limit that all complex phases vanish leads to
\begin{equation}
-Y_{\nu}^T Y_{\nu}=Y_{\nu}^{\dagger}Y_{\nu}\,.
\end{equation}
This is the case considered in \cite{MLFV}, where a close connection is
obtained between experimental low energy data on lepton mixing and the
pattern of various charged LFV processes, that is the
correlation between phenomena A and B in the absence of CP violation.
It corresponds to choosing $R$ and $U_{\nu}$ real.
However, even in this case the RGE between the low energy scale at which
the light neutrino masses and mixings are measured and the scale $M_\nu$
at which $Y_\nu$ is evaluated could have an impact on the correlation 
in question.
\item {\bf Case 2:} Leptonic CP violation at low energies, but no CP violation
  relevant for leptogenesis (barring flavour effects). This corresponds to assuming that the
  leptonic mixing matrix $U_{\nu}$ contains CP violating phases so that 
  $Y_{\nu}^T Y_{\nu}$ is complex, but
  $Y_{\nu}Y_{\nu}^{\dagger}$ is real or equivalently as seen in 
  (\ref{YYdagger}) $R$ is real.
\item {\bf Case 3:} CP violation relevant for leptogenesis but no low energy
  leptonic CP violation. This corresponds to having
  $Y_{\nu}Y_{\nu}^{\dagger}$ and $R$ complex, but $U_{\nu}$ real.
\item {\bf Case 4:} There is leptonic CP violation both at low and high
  energies, that is both $U_{\nu}$ and $R$ are complex quantities. 
 It should be stressed that Case 4 is of course the general case and, 
  in fact, the most
  ``natural'' one, since once CP is violated by the Lagrangian the six
  CP violating phases contained in $m_D$ lead in general to CP
  violation both at low and high energies.
\end{itemize}

\subsection{Final Remarks}
It is clear that (\ref{YdaggerY}), depending on $U_{\nu}$, $R$, $d_\nu$ and 
$M_\nu$ enables one to analyze separately
the four cases considered here.
In each case there will be simultaneously implications for 
lepton
flavour violations, leptogenesis and low energy CP violation and
mixing with certain correlations between them. These correlations can be 
affected by RGE between the low energy scale and $M_\nu$.

At this stage the following comments are in order:
\begin{itemize}
\item
$U_\nu$ is relatively well known from oscillation experiments with the
exception of $s_{13}$, the phase $\delta$ and the Majorana phases $\alpha$ 
and $\beta$. In order to use it for the calculation of $Y_\nu$ it has to be
evolved by RG equations to $M_\nu$.
\item
With the measured two mass differences squared from solar and atmospheric 
oscillation data, the diagonal matrix $d_\nu$ is a function of a single 
parameter that
we choose to be the mass of the lightest neutrino. Again these
parameters have to be evaluated at the scale $M_\nu$ with the help of 
renormalization group techniques.
\item
The matrix $R$ depends on three complex parameters that influence
simultaneously lepton flavour violation and leptogenesis as seen in
(\ref{YdaggerY}) and (\ref{YYdagger}), respectively. Some constraints on $R$ 
can then be obtained from these two phenomena but a complete determination 
of this matrix is only possible in an underlying theory represented usually 
by special texture zeros of $Y_\nu$. 
\item
Finally, $M_\nu$ can be restricted from the BAU in the context of the seesaw
mechanism  and if the eigenvalues 
of the right-handed neutrino matrix $D_R$ are hierarchical,
the absolute lower bound on the lowest $M_i$ is $\ord(10^8)$ or even higher. 
In the case of 
the MLFV considered here the right-handed heavy neutrinos have to be 
quasi-degenerate in order to avoid new flavour violating interactions. 
In this case BAU can be explained with the help of RRL which combines 
the resonant leptogenesis 
considered in \cite{Pilaftsis:1997jf,Pilaftsis:2003gt}
 and radiative leptogenesis  
\cite{GonzalezFelipe:2003fi,Turzynski:2004xy,Branco05}.
The lower bound on $M_\nu$ can be significantly lowered in this case, 
as we will see explicitely below.

\end{itemize}

\boldmath
\section{Radiative corrections in MLFV}\label{sec:RGE}
\unboldmath
\subsection{Preliminaries}
Our MLFV scenario defined in the previous section contains no free parameters
beyond the neutrino masses, the PMNS matrix, a matrix of form $R_{deg}$,
an initial, universal heavy Majorana neutrino mass, and perhaps
additional flavour-blind parameters that depend on the MLFV model.
The rates for charged lepton flavour violation thus follow upon computing
radiative corrections due to the degrees of freedom between the scales
$M_Z$ and $\Lambda_{\rm GUT}$, and with suitable washout factors also
the baryon asymmetry $\eta_B$.

In this section we investigate how the 
CP- and flavour-violating quantities relevant to
leptogenesis and charged lepton flavour violation, respectively,
are radiatively generated.
Since leptogenesis in the present framework can be considered as a
generalization of the setup with two heavy singlets in~\cite{Branco05} to the case of three degenerate flavours, we will also
clarify what novelties arise in this case. This will be important in
comparing our results to the existing literature.

An important point will be that, due to the hierarchy between the
GUT/flavour-breaking scale {\boldmath $\Lambda_{\rm GUT}$} and the neutrino mass scale
$M_\nu$,
large logarithms appear such that the parameter counting for the 
coefficients $c_i$ of flavour structures that has been recently
presented in~\cite{CIP06} should be modified.
Rather than
being independent, the coefficients of structures containing different
powers of Yukawa matrices are related by the renormalization group,
while any additional independent effects are suppressed. Although this fact
in principle increases the predictivity of MLFV, in our phenomenological
sections it will still turn out insufficient to have correlations between
high-scale and weak-scale observables.

\boldmath\subsection{MLFV with a degeneracy scale}
\unboldmath
\label{sec:mlfvdegen}
We have defined our MLFV scenario to have a scale at which the masses
of the right-handed neutrinos are exactly degenerate, such that the
matrix $M_R$ has no flavour structure at all. In general, there will be
additional flavoured particles in the theory. As a specific example, we
consider the MSSM. Here the $N_i$ are accompanied by heavy sneutrinos
$\tilde N^c_i$, and there are also SU(2) doublet sleptons $\tilde l_i$,
transforming as
\be
	\tilde l \to V_L \tilde l,
	\qquad \tilde N^c \to V_{\nu_R}^* \tilde N^c 
\ee
under the transformation~(\ref{eq:trafofields}).
The Lagrangian then contains soft SUSY breaking terms
\be
   {\cal L}_{soft} =-{\tilde N}^{c*}_i {\tilde m}^2_{\nu ij} {\tilde N}^c_j
	- {\tilde l}_i^* {\tilde m}^2_{l i j} {\tilde l}_j  + \dots,
\ee
where the ellipsis denotes further scalar mass matrices and trilinear scalar
interactions.
In general all  matrices in ${\cal L}_{soft}$ have
non-minimal flavour structure.
The simplest generalization of our degenerate scenario is then to extend the
requirement of exact degeneracy to all mass matrices, similar to
minimal supergravity. To be
specific, we require all scalar masses to have the same value $m_0$
at the high scale
and also require the $A$-terms to have the mSUGRA form $A = a Y$
with $Y$ the corresponding Yukawa matrix and $a$ a universal, real parameter
of the theory.
This example also provides us with a concrete value for the scale
$\Lambda_{\rm LFV}$: LFV processes such as $l_i \to l_j \gamma$ are
mediated by loop diagrams involving sleptons and higgsinos or (weak)
gauginos, and unless gaugino
masses are very large, the scalar particles such as $\tilde l_i$ decouple at a
scale $\Lambda \sim m_0$. Hence the operators governing
charged LFV are suppressed by powers of $m_0 \equiv \Lambda_{\rm LFV}$.
As in the case of the heavy Majorana masses, the generalized
degeneracy requirement is not stable under radiative corrections,
and for the same reason it is not renormalization scheme independent.

\boldmath\subsection{Radiatively generated flavour structure and
large logarithms}
\unboldmath
\label{sec:largelog}

As will be discussed in detail in the following section, the CP asymmetries
necessary for leptogenesis require mass splittings between the decaying
particles. The decaying particles are on their mass shell\footnote{
We follow the treatment
of~\cite{Pilaftsis:1997jf,Pilaftsis:2003gt} (see also
\cite{D'Ambrosio:2003wy}), where sometimes the on-shell masses are
replaced by thermal masses. (We will employ zero-temperature masses.)
}, but 
the degenerate initial conditions are usually specified in a massless
scheme\footnote{This is likely appropriate if the degeneracy is true
to some flavour symmetry of an underlying theory, relating high-energy
Lagrangian parameters and broken at the scale $\Lambda_{\rm GUT}$.}
(\barms\ to be definite~\cite{Bardeen:1978yd}).

At one loop, the two mass definitions are related by a formula of the structure
\be
	M_i^{os} = M_i^{\mbox{\tiny\barms}}(\mu)
		   + c_i M_i^{\mbox{\tiny\barms}}(\mu) \ln \frac{M_i}{\mu}
			+ \mbox{nonlogarithmic corrections} ,
		\label{eq:onshell}
\ee
where $\mu \sim \Lambda_\mathrm{GUT}$ is the \barms\ renormalization scale,
$c_i = 2 (Y_\nu Y_\nu^\dagger)_{ii}/(16 \pi^2)$ in the standard-model seesaw,
and the nonlogarithmic corrections depend on our choice of massless
(or any other) renormalization scheme.
The resulting scheme dependence cannot be present in physical
observables such as the BAU. Since this issue is usually not discussed
in the literature on lepton flavour violation, let us elaborate on
how it may be resolved.

First, notice that while the nonlogarithmic terms
in~(\ref{eq:onshell}) are scheme dependent, the logarithmic
corrections proportional to $c_i$ are actually scheme independent.
If $\ln \Lambda_\mathrm{GUT}/M_\nu \gg 1$, the logarithmic terms
must be considered ${\cal O}(1)$ and summed to all orders. This is achieved
in practice by solving renormalization group equations. Similar resummations
must be performed for all other parameters in the theory (such as
Yukawa couplings).
Correspondingly, the dominant
higher-loop corrections to LFV observables and leptogenesis are
approximated by using leading-order expressions with one-loop
RGE-improved Yukawa couplings and masses. This is the
leading-logarithmic approximation. Nonlogarithmic corrections such as
those indicated in~(\ref{eq:onshell}) are then sub-leading and should
be dropped.

What happens when the logarithms are not large is the following. If the
MLFV framework is an effective theory for some fundamental theory
where the degeneracy is enforced by a flavour symmetry, for instance
the group~(\ref{eq:flavourgroup}), then the
degeneracy holds in {\em any} scheme (that respects the symmetry)
in the full theory and the scheme
dependence observed in~(\ref{eq:onshell}) must be due to unknown
threshold corrections in matching the underlying and effective theories.
Since the flavour symmetry in MLFV, by definition,
is broken precisely by the Yukawa matrices,
this matching introduces all possible terms that are invariant under
transformations
(\ref{eq:trafofields},\ref{trafo1},\ref{trafo2},\ref{trafo3}).
A list of such structures has
recently been given in~\cite{CIP06}, for instance,
\be					\label{eq:MRparam}
  M_R = M_\nu \Big[ 1 + c_1 (Y_\nu Y_\nu^\dagger + (Y_\nu Y_\nu^\dagger)^T)
	+ c_2 (Y_\nu Y_\nu^\dagger Y_\nu Y_\nu^\dagger
		+ (Y_\nu Y_\nu^\dagger Y_\nu Y_\nu^\dagger)^T) + \dots
	\Big] .
\ee
The coefficients $c_1$ and $c_2$ have been claimed by these authors
to be independent ${\cal O}(1)$ coefficients. Indeed  these terms contain only
non-logarithmic terms and (small) decoupling logs when $M_R$
is taken in the \barms\ scheme, renormalized near the GUT (matching) scale.

However, when computing the (physically relevant) on-shell $M_R$
in the case of $\Lambda_{\rm GUT} \gg M_\nu$, large logarithms
dominate both $c_1$ and $c_2$.
The leading logarithmic contributions are not independent, but
are related by the renormalization group. $c_2$ is quadratic in
$L \equiv \ln \Lambda_\mathrm{GUT}/M_\nu$, while $c_1$ is linear, and
the RGE for $M_R$ implies $c_2|_{L^2} = \frac{1}{2} [c_1|_{L}]^2$.
These logs are summed by RG-evolving $M_R^{\overline{\rm MS}}$ to a scale
$\mu \sim M_\nu$. The additional conversion to on-shell masses is
then again a sub-leading correction.

Finally, we note that if there is no underlying symmetry,
the degeneracy condition can again be true at most for special choices of
scheme/scale, and must be fine-tuned.

Numerically, the logarithms dominate already
for mild hierarchies $\Lambda_\mathrm{GUT}/M_\nu > 10^2$, as then
$2 \ln\Lambda_\mathrm{GUT}/M_\nu \approx 10$. Let us now
restrict ourselves to hierarchies of at least two orders of magnitude and
work consistently in the leading-logarithmic approximation. As explained
above, in this case non-logarithmic corrections both of the threshold
type (in the coefficients $c_i$ in~(\ref{eq:MRparam}) and
in physical quantities
(on-shell masses, CP asymmetries, etc.) are sub-leading and should be dropped.
In this regard our apparently ``special'' framework of initially
degenerate heavy neutrinos turns out to be the correct choice at
leading-logarithmic order.

Finally we recall that the positions of the poles of the $N_i$
two-point functions contain an imaginary part related to the widths of
these particles.
While not logarithmically enhanced, these are also scheme-independent
at one loop (as the widths are physical), and it is unambiguous to include them
in applications. In fact, these widths effects are often numerically
important for the CP asymmetries in $N_i$
decay~\cite{Pilaftsis:1997jf,Pilaftsis:2003gt},
and we will keep them in our numerical analysis.

\subsection{Renormalization-group evolution: high scales}
\label{sec:RGE-high}
For the running above the seesaw scale the relevant
renormalization-group equations have been given in
in~\cite{Antusch:2002rr} (in particular, last paper)
for the SM and MSSM seesaw models.
As the physical quantities studied below,
such as leptonic CP asymmetries, involve mass eigenstates,
it is convenient to keep the singlet mass matrix
diagonal during evolution (see, e.g.,~Appendix B
of~\cite{Antusch:2003kp}):
$$
	M_R(\mu) = \mathrm{diag}(M_1(\mu), M_2(\mu), M_3(\mu)) .
$$
Defining 
\be
H=Y_{\nu}Y_{\nu}^{\dagger},
\ee
and
\be
t = \frac{1}{16 \pi^2} \ln\left( \mu/\Lambda_{\rm GUT}\right),
\ee
one obtains for the mass eigenvalues in the SM with right handed neutrinos:
\be					\label{RGErightmasses}
\frac{dM_i}{dt} = 2 H_{ii}\,M_i \qquad \mbox{(no sum)} .
\ee
Note that due to the positivity of the right-hand side
of~(\ref{RGErightmasses}),
the running will always decrease the masses when running from the GUT
to the seesaw scale.

The matrix $H$
satisfies the RGEs
\bea					\label{RGEneutrinoYukawa}
\frac{dH}{dt} &=& [T, H] + 3\, H^2 - 3 \,Y_\nu Y_E^\dagger Y_E Y_\nu^\dagger
+ 2 \alpha H              \qquad  \mbox{(SM)},  \label{eq:HSM} \\
\frac{dH}{dt} &=& [T, H] + 6\, H^2 + 2\, Y_\nu Y_E^\dagger Y_E Y_\nu^\dagger
+ 2 \alpha H               \qquad  \mbox{(MSSM)}  \label{eq:HMSSM},
\eea
where
\bea
 \alpha &=& Tr(Y_{\nu}^{\dagger} Y_{\nu})+Tr(Y_{e}^{\dagger}Y_{e})
	+3\, Tr(Y_{u}^{\dagger}Y_{u})+3\, Tr(Y_{d}^{\dagger}Y_{d})
	-\frac{9}{20}g_1^2-\frac{9}{4}g_2^2 \quad \mbox{(SM)}, \\
 \alpha &=& Tr(Y_{\nu}^{\dagger} Y_{\nu})
	+3\, Tr(Y_{u}^{\dagger}Y_{u})
	-\frac{3}{5}g_1^2-3 g_2^2 \quad \mbox{(MSSM)}, \\
								\label{Tdef}
\!\! T_{ij} &=& \left\{ \begin{array}{ll}
		- \frac{M_j +M_i}{M_j-M_i} \mathrm{Re} H_{ij}
		-i \frac{M_j-M_i}{M_j+M_i}\mathrm{Im}H_{ij} & (i \not=
                j, \quad \mbox{SM}),
  \\ 
		- 2\, \frac{M_j +M_i}{M_j-M_i} \mathrm{Re} H_{ij}
		- 2\, i\, \frac{M_j-M_i}{M_j+M_i}\mathrm{Im}H_{ij} & (i \not=
                j, \quad \mbox{MSSM}),
		\\ 0 & (i = j) , \end{array} \right.
\eea
and GUT normalization has been employed for $g_1$.
The matrix $T$ satisfies $\dot U = T U$, where
$M_R^{(0)}(\mu) = U(\mu)^T M_R(\mu) U(\mu)$ and $M_R^{(0)}$ satisfies
the unconstrained RGEs given in~\cite{Antusch:2002rr}.
Note that
$\alpha$ is real and has trivial flavour structure. Note the different
relative signs in~(\ref{eq:HSM}) and~(\ref{eq:HMSSM}); we will return
to this point below.

We now turn to a qualitative analysis of these equations and their
impact on leptogenesis and flavour violation.
Ignoring flavour effects in the Boltzmann evolution of charged leptons,
the baryon asymmetry $\eta_B$ is approximately proportional to the combinations
${\rm Im} ((H_{ij})^2) = 2\, {\rm Re} H_{ij}\, {\rm Im} H_{ij}$ $(i \not= j)$,
evaluated in the mass eigenbasis.
At the scale $\Lambda_{\rm GUT}$, degeneracy of $M_R$ allows the use of an
$SO(3)$ transformation to
make the off-diagonal elements of  ${\rm Re} H$ vanish.\footnote{
To see this, 
notice that $H$ is hermitian, so $\RE H$ is real symmetric. That is, it
can be diagonalized by a real orthogonal (and hence unitary) transformation
of the right-handed neutrinos. Now if all three neutrinos are degenerate,
such a rotation affects no term in the Lagrangian besides $Y_\nu$.
}
As explained above, we should RG-evolve all parameters to the
scale $\mu \sim M_\nu$ to avoid large logarithms. Let us first consider
the formal limit of vanishing charged lepton Yukawa couplings $Y_E$
for the SM case.
It is instructive to split~(\ref{eq:HSM}) into real and imaginary parts.
The former satisfies
\be		\label{eq:heq}
	\frac{d {\RE} H}{dt} = [\RE T, \RE H] - [\IM T, \IM H] +
		3 \Big\{ (\RE H)^2 - (\IM H)^2 \Big\} + 2 \alpha \RE H.
\ee

To investigate how a nondiagonal $\RE H$ can be generated radiatively, assume that
it is zero at some scale (initial or lower). Then~(\ref{eq:heq})
reduces to
\be		\label{eq:heq0}
	\frac{d \RE H}{dt} = - [\IM T, \IM H] - 3 (\IM H)^2 .
\ee
(At $t=0$, an extra term proportional to the
  offdiagonal part of $(\IM H)^2$ appears on the right-hand side
of~(\ref{eq:heq0}).)
Now evaluate this for the $(2,1)$ element and notice that
$T_{ij}=0$ and $\IM H_{ij}=0$ for $i=j$.
If there were only two heavy singlets in the theory,
each term in each matrix product would require one $(2,1)$ element
and one $(1,1)$ or $(2,2)$ element from the two matrix factors. For example,
\be
	(\IM T\, \IM H)_{21} = \IM T_{21} \underbrace{\IM H_{11}}_0
					+ \underbrace{\IM T_{22}}_0 \IM H_{21}
			= 0,
\ee
and similarly for the other terms.
Consequently,
\be
	\RE H_{21} = 0 \Rightarrow \frac{d \RE H_{21}}{d t} = 0 .
\ee
We see that there is no radiative leptogenesis
in the two-flavour case when $Y_E=0$.
This is consistent with the approximate equation (12) in~\cite{Branco05},
where $\RE H_{21}$ was found to be proportional to $y_\tau^2$.
It is easy to see that the argument breaks down in the three-flavour case.
For instance,
\be
	((\IM H)^2)_{21} = \IM H_{21} \IM H_{11}
			+ \IM H_{22} \IM H_{21} + \IM H_{23} \IM H_{31}
		= \IM  H_{23} \IM H_{31} ,
\ee
which is in general not zero. The other terms in~(\ref{eq:heq0}) are also
proportional to $\IM  H_{23} \IM H_{31}$. We see that three generations
of heavy neutrinos are necessary and sufficient to generate leptogenesis
without help from charged lepton Yukawas.

Once we restore the charged lepton Yukawas, they will also
contribute. The important qualitative difference is that, whereas
the contribution involving the charged-lepton Yukawas is only logarithmically
dependent of the seesaw scale (as seen in eqs.(\ref{epsigen})--(\ref{Dj})
below for the two-flavour case, or from~\cite{Pilaftsis:2003gt} for the
three-flavour case),
the pure $Y_\nu$ contribution to the radiatively generated $\RE H_{ij}$
scales with $M_\nu$ because it contains two extra powers of $Y_\nu$ as observed in the three flavour scenario studied in \cite{CIP06}.

In summary, we expect the following qualitative behavior for the
BAU as a function of $M_\nu$:
\begin{itemize}
  \item  For small $Y_\nu$ (small $M_\nu$), the dominant contribution to
  $\RE H_{ij}$  and hence to $\eta_B$ should be due to $Y_E$. $\eta_B$ turns
  out to be weakly dependent on $M_\nu$.
  \item For large $Y_\nu$ (large $M_\nu$), in the three-flavour case there
  is a relevant contribution proportional to $((\IM H)^2)_{ij}$. Since
  it contains two extra powers of $Y_\nu$ with respect to the contribution
  proportional to $y_\tau^2$, $\eta_B$ scales linearly with $M_\nu$.
  \item In the case of only two heavy flavours,
  $\eta_B$ is weakly dependent on $M_\nu$ over the whole range of $M_\nu$.
  We will therefore include an ``effective'' two-flavour scenario in our
  numerical analysis.
\end{itemize}
Let us stress that we reached these qualitative conclusions
only upon neglecting flavour effects in the Boltzmann evolution
of the products of the $N_i$ decays.
We will return to these points in Section~\ref{sec:numerics} and
in Section~\ref{sec:leptogenesis}, where we perform a detailed
quantitative analysis.

Finally, let us briefly discuss $l_i \to l_j \gamma$.
In MLFV these radiative lepton decays are governed by
$\Delta_{ij} \equiv Y_\nu^\dagger Y_\nu$ (and structures
involving more powers of Yukawa matrices).
In the case of the SM,
the rates are known to be essentially zero due to a near perfect
GIM cancellation among the tiny neutrino masses. From the point of
view of MLFV, this smallness can be
traced to the fact that, in the SM,
the LFV scale is equal to the LNV scale $\sim M_\nu$.

On the other hand, in the more generic case of the MSSM,
there are additional contributions
mediated by slepton-higgsino or slepton-gaugino loops suppressed
only by a scale $\Lambda_{\rm LFV} \sim m_{\tilde l}$, of order TeV,
as discussed in Section~\ref{sec:mlfvdegen}.
Linearizing the RG evolution, the charged slepton soft mass matrix
acquires the form\cite{Borzumati:1986qx}
\be                \label{eq:sleptonflav}
	\tilde m^2_l (M_\nu) =
		m_0^2 {\bf 1}
		 - L \frac{Y_\nu^\dagger Y_\nu}{16 \pi^2}
			(6 m_0^2 + 2 a_0^2)+ \dots ,
\ee
where the dots denote terms governed by charged lepton Yukawa couplings
$Y_E$ or conserving lepton flavour.
Note that the flavour structure in the soft terms is generated at a high
scale and that, unlike the case of CP asymmetries in $N_i$ decay, the
necessary flavour structure $\Delta$ is already present at the initial
scale $\Lambda_{\rm GUT}$. Hence the RGE running of $\Delta$ merely gives
a correction. Note also that there is dependence on the MLFV model beyond the
choice of LFV scale due to the (in general unknown) RGE coefficients
in (the relevant analog of)~(\ref{eq:sleptonflav}).

\boldmath
\subsection{RGE evolution below $M_\nu$: PMNS matrix and $\Delta_{ij}$}
\unboldmath

So far we have ignored renormalization effects in equations such
as~(\ref{yukawaparamet}), identifying $d_\nu$ and $U_\nu$ with the
physical (light)
neutrino masses and mixing matrix, while the objects $Y_\nu$ and $D_R$
are defined at a high scale.
However, to be orthogonal the matrix $R$ has to be defined with all
objects given at the same scale.
Now it is well known that using low-energy inputs in $d_\nu$
can be a bad approximation because there are significant radiative
corrections between the weak and GUT scales. However, as investigated in
\cite{Antusch:2003kp}, both in the
SM and in the MSSM with small $\tan \beta$ the main effect
below $M_\nu$ is an approximately universal rescaling of the light
neutrino masses.
This results in larger magnitudes of the elements of $Y_\nu$ extracted
by means of~(\ref{yukawaparamet}) but in a weak running of the matrix
$U_\nu$.
Above the scale $M_\nu$, even though
the heavy singlets are now dynamical, one can still {\em define}
an effective neutrino mass matrix through the seesaw
relation~(\ref{eq:mnueff}). However, the evolution becomes
more involved, as in the presence of heavy singlets there are
additional contributions to the running involving $Y_\nu$. To deal with
this situation, where some of our inputs are specified at the weak scale,
while the matrix $R_{\deg}$ is defined at the scale $\Lambda_\mathrm{GUT}$,
we employ an iterative procedure detailed in Appendix~\ref{app:iterative}.
As was the case for the evolution above $M_\nu$, also
the RGE effects below $M_\nu$, and consequently the relation of
e.g $Y_\nu^\dagger Y_\nu$ to the input parameters necessarily depends
on the details of the MLFV model.

\boldmath
\section{Numerical Analysis: $B(l_i \to l_j \gamma)$ and
CP asymmetries in $\nu_R$ decay}
\unboldmath
\label{sec:numerics}
\subsection{Preliminaries}
For our numerical analysis we take our input parameters at the
weak scale, except for the matrix $R_{\deg}$, which has to be defined
at the scale $\Lambda_{\rm GUT}$. From these inputs we find a
consistent set of parameters at the seesaw scale $M_\nu$,
where the CP asymmetries as well as $B(l_i\to l_j \gamma)$ are
calculated, through the iterative procedure given in
Appendix~\ref{app:iterative}. For the running we use
the package REAP~\cite{Antusch:2005gp}.\\
For the PMNS matrix we use the convention:
\be\label{standard}
U_\nu=
\left( 
\begin{array}{ccc}
c_{12}c_{13} &s_{12}c_{13} &s_{13}e^{-i\delta} \\
-s_{12}c_{23}-c_{12}s_{23}s_{13}e^{i\delta} &c_{12}c_{23}-s_{12}s_{23}s_{13}e^{i\delta}  & s_{23}c_{13} \\
s_{12}s_{23}-c_{12}c_{23}s_{13}e^{i\delta} & -s_{23}c_{12}-s_{12}c_{23}s_{13}e^{i\delta}  & c_{23}c_{13} 
\end{array}
\right)\cdot V
\ee
and $V={\rm Diag}(e^{i\alpha/2}, e^{i\beta/2},1)$ where $\alpha$ and
$\beta$ denote the Majorana phases and $\delta$ denotes the Dirac phase.
We parameterize the complex orthogonal matrix $R$ as follows:
\be\label{RMatrix}
R=
\left( 
\begin{array}{ccc}
\hat c_{12}&\hat s_{12}&0\\
-\hat s_{12}&\hat c_{12}&0 \\
 0&0&1
\end{array}
\right)
\left( 
\begin{array}{ccc}
1&0&0\\
0&\hat c_{23}&\hat s_{23}\\
0&-\hat s_{23}&\hat c_{23}
\end{array}
\right)
\left( 
\begin{array}{ccc}
\hat c_{13}&0&\hat s_{13}\\
0&1&0\\
-\hat s_{13}&0&\hat c_{13}
\end{array}
\right),
\ee
with $\hat s_{ij}\equiv \sin{\hat \theta_{ij}}$, with $\hat
\theta_{ij}$ in general complex:
\begin{equation}\label{thetahat}
\hat \theta_{ij}=x_{ij}+i\,y_{ij}.
\end{equation}
In the degenerate case, the angles $x_{ij}$ can be made to vanish by
a redefinition of the right-handed neutrinos, i.e.\ a matrix of the
form $R_{deg}$ is parameterized by three real numbers $y_{ij}$.

In the following, we use maximal atmospheric mixing
$c_{23}=s_{23}=1/\sqrt{2}$ and a solar mixing angle
 $\theta_{\rm sol}=33^\circ$, with corresponding values for its
sine $s \equiv s_{12}$ and cosine $c \equiv c_{12}$.
For the sine of the CHOOZ angle $s_{13}$ and the phases we allow the ranges
\be\label{In0}
\qquad  0\le s_{13}\le 0.25,\qquad 0<\alpha, \beta, \delta <2\pi,
\ee
and for the light neutrinos we use the low energy values 
\be\label{In1}
\Delta m^2_{\rm sol}=m_{ 2}^2-m^2_{ 1}=8.0\cdot 10^{-5}~\ev^2
\ee
\be\label{In2}
\Delta m^2_{\rm atm}=| m_{3}^2-m^2_{ 2}|=2.5\cdot 10^{-3}~\ev^2
\ee
\be\label{In3}
0\le m_\nu^{\rm lightest}\le 0.2~{\rm eV}
\ee
with $m_\nu^{\rm lightest}=m_{1}(m_{3})$ for normal (inverted) hierarchy,
respectively.
See \cite{Strumia:2005tc} for a detailed discussion of the neutrino
masses and mixing. For the heavy neutrino mass scale,
we consider a wide range
\be
10^6\, {\rm GeV}<M_\nu<10^{14}\, {\rm GeV},
\ee 
and the CP violating parameters $y_{ij}$ are all taken in the
range $[-1,1]$ if not otherwise stated.

\boldmath\subsection{Perturbativity bounds}
\unboldmath
In the MLFV framework the magnitudes of the Yukawa couplings $Y_{\nu}$ are
very sensitive
to the choice of $M_\nu$, $m_\nu^\mathrm{lightest}$
and the angles in the matrix $R_\mathrm{deg}$, as is evident
from (\ref{yukawaparamet}). To render the framework
perturbative, we impose the constraint
\begin{align}
  \frac{y_\mathrm{max}^2}{4 \pi} < 0.3 ,
\end{align}
where $y_\mathrm{max}^2$ is the largest eigenvalue of $Y_\nu^\dagger Y_\nu$ .
By means of (\ref{YdaggerY}), it translates into a bound on $R^\dagger R=R^2$
and the angles $y_{ij}$ that scales with $M_\nu^{-1}$ and hence is most severe
for a large lepton-number-violating scale. Analogous bounds apply to
other dimensionless couplings whose number depends on the precise MLFV
model. For instance, in the SM there is also the Higgs self coupling
$\lambda_H$, whereas in the MSSM there is no such additional coupling.

\subsection{ Lepton Flavour Violation and $l_i \to l_j \gamma$}
%
Following Cirigliano {\it et al.} \cite{MLFV}
 we consider the normalized branching
fractions defined as
\begin{equation}
B(l_i\to l_j \gamma)=\frac{\Gamma(l_i\to l_j \gamma)}{\Gamma(l_i\to
  l_j\nu_i \bar \nu_j)}\equiv r_{ij} \hat B(l_i\to l_j \gamma),
\end{equation}
where $\hat B(l_i\to l_j \gamma)$ is the true branching ratio and
$r_{\mu e}=1.0$, $r_{\tau e}=5.61$ and $r_{\tau\mu}=5.76$.
Assuming first the heavy right-handed neutrinos to be degenerate but
not making the assumptions of $R=\mathbbm{1}$ and $U_\nu$ being real as done in
\cite{MLFV}, the straightforward generalization of (29) in \cite{MLFV} is
\begin{equation}\label{Branchingratio}
B(l_i\to l_j \gamma)=
384 \pi^2 e^2 \frac{v^4}{\Lambda_{\rm LFV}^4}|\Delta_{ij}|^2|C|^2.
\end{equation}

Here $v=174$ GeV is the vacuum expectation value of the SM Higgs
doublet,\footnote{$v = \sqrt{v_1^2 + v_2^2}$ for two-Higgs-doublet models such
as the MSSM. Powers of $\sin\beta$ can be absorbed into $C$ or into
a redefinition $\Lambda_{\rm LFV} \to \Lambda_{\rm LFV}^{\rm eff}$}
$\Lambda_{\rm LFV}$ is the scale of charged lepton flavour violation,
and $C$ summarizes the Wilson coefficients of the relevant operators
that can be calculated in a given specific model. They are naturally
of ${\cal O}(1)$ but can be different in different MLFV models.
As we would like 
to keep our presentation as simple as possible, we will set $|C|=1$  in what
follows, bearing in mind that in certain scenarios
$C$ may differ significantly from unity.
Thus the true $B(l_i\to l_j \gamma)$ can be different from our estimate in a
given MLFV model, but as $C$ is, within MLFV, independent of external
lepton flavours, the ratios of branching ratios take a very simple form
\begin{equation}\label{Deltaratio}
\frac{B(l_i\to l_j \gamma)}{B(l_m\to l_n \gamma)}=
\frac{|\Delta_{ij}|^2}{|\Delta_{mn}|^2}.
\end{equation}

The most important objects in (\ref{Branchingratio}) and
(\ref{Deltaratio}) are
\begin{equation}\label{Deltaij}
\Delta_{ij}\equiv (Y_\nu^\dagger Y_\nu)_{ij}=\frac{1}{v^2} (U_\nu
\sqrt{d_\nu}R^\dagger D_R R \sqrt{d_\nu}U_\nu^\dagger)_{ij},
\end{equation}

which in the limit of $R=\mathbbm{1}$, $D_R=M_\nu\mathbbm{1}$,
and $U_\nu$ being real reduce to
$\Delta_{ij}$ as given in (14) of \cite{MLFV}.

With the formula (\ref{Deltaij}) at hand we can generalize the
expressions for $\Delta_{ij}$ in (24) of \cite{MLFV} to the general case
of complex $R$ and $U_\nu$. 
To this end we will use the standard
parametrization of the PMNS matrix $U_\nu$ in~(\ref{standard}) and the parametrization of $R$ in~(\ref{RMatrix}).
As the general expressions for $\Delta_{ij}$ in terms of $x_{ij}$
and $y_{ij}$ are very complicated, we give in Appendix B explicit formulae 
setting all $x_{ij}=y_{ij}=0$ except for $y_{12}\neq 0$. We will
discuss in our numerical analysis also 
the cases for which  $y_{13}$ and $y_{23}$ are
non-vanishing.
As mentioned above, setting 
$x_{ij}=0$ is in accord with the degeneracy of the right-handed neutrinos. 
Once this degeneracy is broken by RG effects, the $x_{ij}$ become non-zero.

Recall from Section~\ref{sec:RGE} that $\Delta_{ij}$ evolves above
the scale $M_\nu$ and the flavour structures it affects, such as the
slepton mass matrix $m^2_{\tilde l}$, also evolve between $M_\nu$ and
$\Lambda_{\rm LFV}$ (and the resulting effective operators below
$\Lambda_{\rm LFV}$ also evolve). Moreover, the flavour-violating
piece in, for example, $m^2_{\tilde l}$ is not exactly proportional
to $\Delta$ at the scale $M_\nu$ beyond leading order
because these objects satisfy different RGEs between $M_\nu$ and
$\Lambda_{\rm GUT}$.
All this running depends, beyond the operator, also on the
details of the model. Below the seesaw scale the flavour-non-universal
contributions are governed by $Y_E$ (although trilinear couplings such
as the $A$-terms in the MSSM can also contribute), which is
analogous to the case of the PMNS matrix.
Based on the experience that the running of the PMNS angles is weak
in the SM and the MSSM unless $\tan\beta$ (and hence $y_\tau$) is large, we
ignore all these details and evaluate $\Delta_{ij}$ at the scale
$M_\nu$.

That $\Delta_{ij}$ has to be evaluated at the high energy scale
$M_\nu$, and hence $U_\nu$ and $d_\nu$ 
have to be evaluated at $M_\nu$ by means of renormalization group equations 
with the initial conditions given by their values at $M_Z$,
has recently been stressed in particular in \cite{Petcov:2005jh}.
The dominant contributions to the flavour-violating pieces in the
charged slepton masses matrix in the MSSM that is relevant for 
$l_i \to l_j\gamma$ are proportional to $Y_\nu^\dagger Y_\nu$ and
come from scales above $M_\nu$, as seen for instance in  equation (30)
of \cite{Deppisch:2005rv} (where charged lepton Yukawas and $A$-terms
have been dropped and only contribute at higher orders)
and the fact that right-handed neutrinos and their Yukawa
couplings are absent below that scale.

All other parameters of a given MLFV model,
hidden in the 
Wilson coefficient $C$ in (\ref{Branchingratio}), like slepton and chargino
masses in the MSSM, would have to be evaluated at the electroweak
scale and lower scales if a concrete value for $C$ was desired. 

\begin{figure}[tb]
\vspace{0.01in}
\centerline{
\epsfxsize=0.55\textwidth
\epsffile{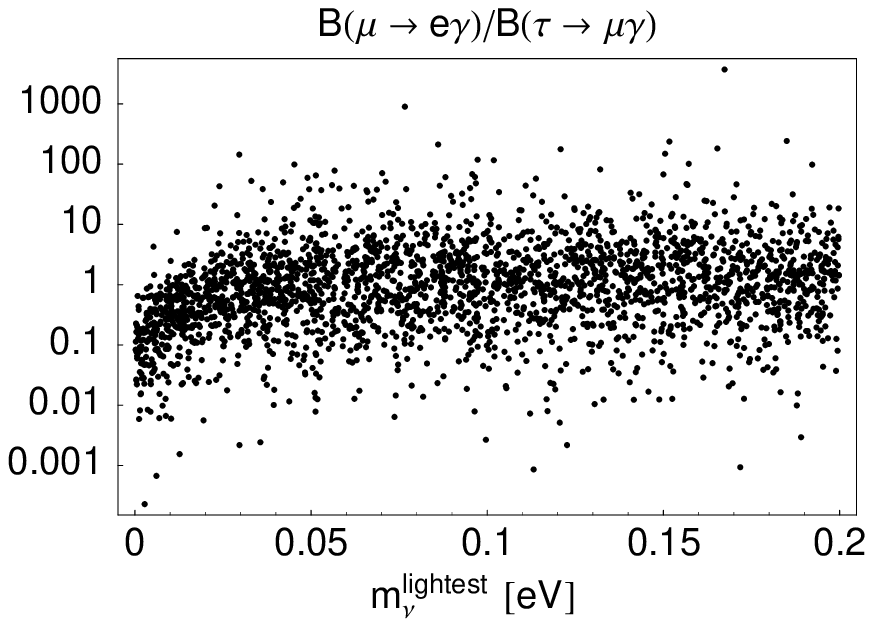}
\hspace{-0.3in}
\epsfxsize=0.55\textwidth
\epsffile{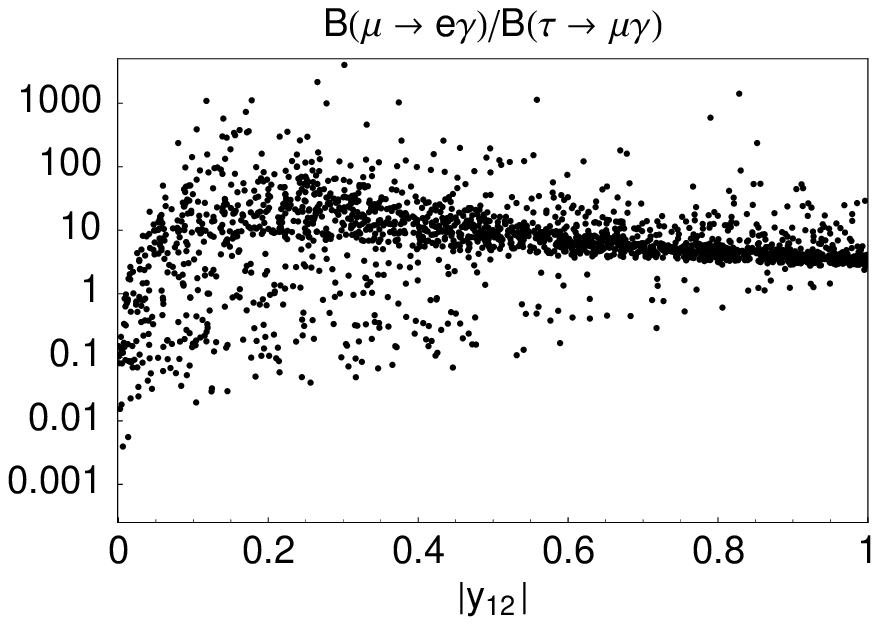}
}
\caption{Double ratios of $l_i \to l_j \gamma$ for the MSSM with
$\tan\beta=2$. Left plot: All parameters varied, right plot: no phases
and only $y_{12} \not= 0$. For a discussion, see the text.
\label{fig:LFVdoubleratio} }
\end{figure}
The ratio $B(\mu \to e \gamma)/B(\tau \to \mu \gamma)$ is shown for
the case of the MSSM with $\tan\beta=2$ in
Fig.~\ref{fig:LFVdoubleratio} (left). All other parameters are
varied in the ranges given above.
We see that this ratio varies over about six orders of magnitude and
$B(\mu \to e \gamma)$ can be a factor $10^3$ larger than $B(\tau \to \mu
\gamma)$ in qualitative agreement
with~\cite{Kanemura:2005cq,Petcov:2006bn}. We have checked that the leptogenesis constraint, as discussed in Section~\ref{sec:leptogenesis}, has no significant impact.
This contradicts the findings of~\cite{CIP06}.
Even when constraining the Dirac and Majorana phases in the PMNS
matrix to zero
and allowing only for a single non-vanishing angle $y_{12}$ at the
scale $\Lambda_{\rm GUT}$, we can still have
$B(\mu \to e \gamma) \gg B(\tau \to \mu\gamma)$. This is again in
agreement with~\cite{Kanemura:2005cq,Petcov:2006bn}.
We will consider the single ratio $B(\mu \to e \gamma)$ together with
the leptogenesis constraint in Section~\ref{sec:leptogenesis}.

\begin{figure}[tb]
\vspace{0.01in}
\centerline{
\epsfxsize=0.6\textwidth
\epsffile{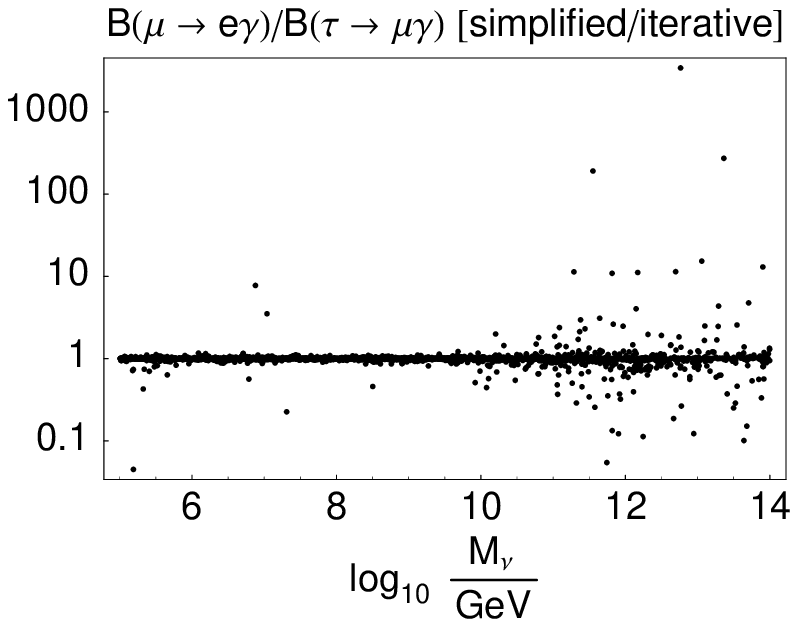}
\hspace{-0.5in}
\epsfxsize=0.6\textwidth
\epsffile{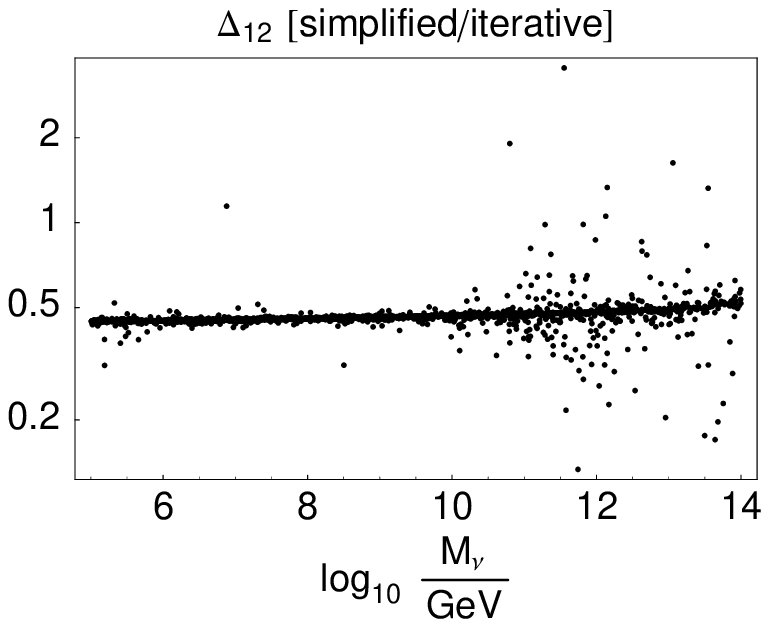}
}
\caption{Impact of iterative vs simplified procedure. Left plot:
Simplified result for the ratio of branching
ratios $B(\mu \to e \gamma)/B(\tau \to \mu
\gamma)$, normalized to the one obtained with the
iterative procedure. Right plot: Similarly for $\Delta_{12}$.
\label{fig:LFVcomparison} }
\end{figure}
It is also interesting
to compare our elaborate iterative procedure of matching high- and
low-energy parameters to a simpler procedure where we simply impose the
weak-scale PMNS and neutrino mass parameters at the scale
$\Lambda_{\rm GUT}$ (Fig.~\ref{fig:LFVcomparison} (left), corresponding to
the MSSM with $\tan\beta=2$). It turns out that
both procedures agree well for small scales $M_\nu$. (This agreement
is slightly worse for $\tan\beta=10$.) For large
values $M_\nu > 10^{11}$ GeV, deviations up to a few orders of magnitude can
occur for some choices of parameters. It appears that this is usually
due accidentally small branching ratios in one of the approaches. This
is supported by the right plot in the Fig.~\ref{fig:LFVcomparison}, which shows a good
agreement for the more fundamental flavour-violating quantity  $\Delta_{12}$
up to the (expected) different overall normalization.

\subsection{CP asymmetries}
\label{sec:CPA-numerics}
\begin{figure}[h]
\vspace{0.01in}
\centerline{
\epsfxsize=0.54\textwidth
\epsffile{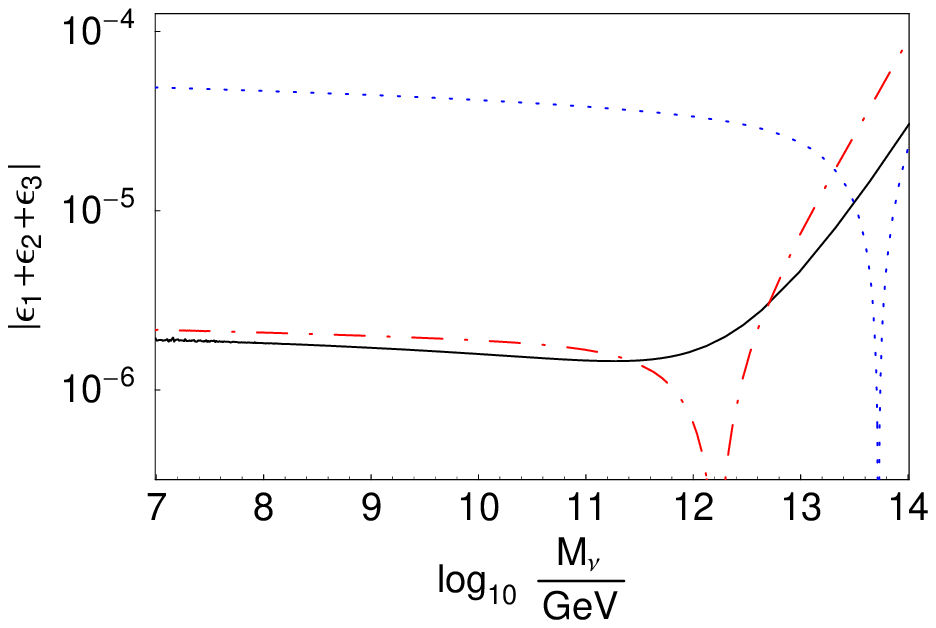}
\hspace{-0.3in}
\epsfxsize=0.54\textwidth
\epsffile{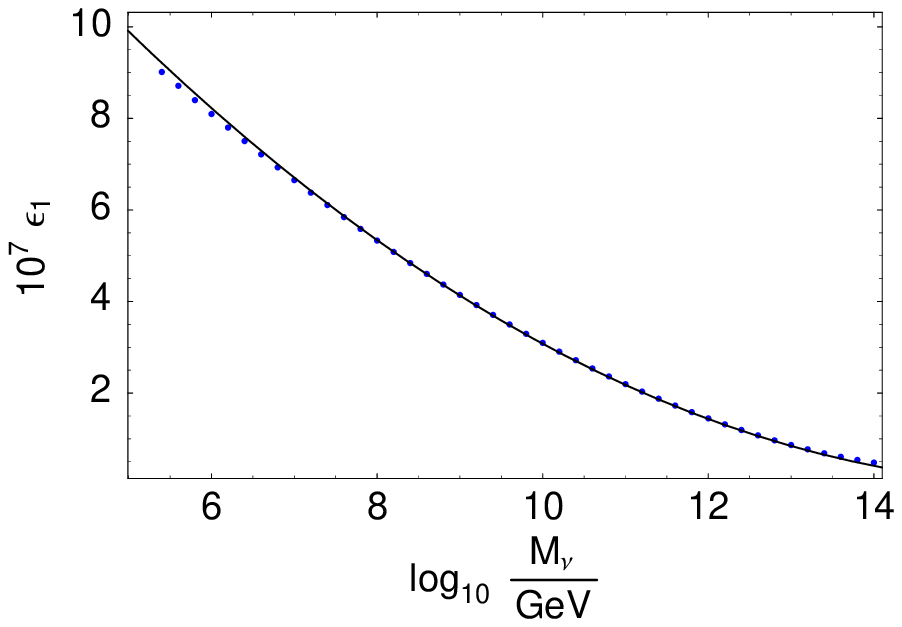}
}
\caption{Left plot: $M_\nu$ dependence of $| \sum_i \epsilon_i|$
for the generic (3-flavour) case. Right plot: effective 2-flavour case.
Normal hierarchy, $m_\nu^\mathrm{lightest}=0.02$ eV ;
$y_{12}=0.8$, $y_{13}=0.2$, $y_{23}=0.6$ (3-flavour case), $y_{12}=1$
and $y_{13}=y_{23}=0$
(effective 2-flavour case). The PMNS phases have been taken to
be $\delta = \alpha = \beta = \pi/10$.
Right plot: Effective two-flavour case; only $\epsilon_1$ is shown, on
a linear scale.
\label{fig:sect4eps1} }
\end{figure}
We are also in a position to illustrate and check numerically our
qualitative discussion in Section~\ref{sec:RGE} of the CP asymmetries
relevant for leptogenesis. A thorough investigation of the baryon asymmetry
follows in the next section.
Fig.~\ref{fig:sect4eps1} shows the sum of the three CP
asymmetries $| \sum_i \epsilon_i |$ defined below~(\ref{eq:cpas}),
for the generic three-flavour case (left plot) and the
CP asymmetry $\epsilon_1$ for the effective two-flavour case
where only $y_{12}\not=0$ (right plot).
One can see clearly that in the latter case the
dependence on $M_\nu$ is weak and slightly reciprocal. In fact
this dependence is approximately
proportional to $\ln^2\Lambda_{\rm GUT}/M_\nu$ (black solid line) in
agreement with expectations.
The generic case is shown in the left plot for the SM (black solid) as well as
the MSSM for $\tan\beta=2$ (red dot-dashed) and $\tan\beta=10$ (blue
dotted), with the
remaining parameters given in the Figure caption.
In contrast to the two-flavour case, there is strong
dependence on $M_\nu$ for $M_\nu > 10^{12}$ GeV, when the contribution
due to $Y_\nu$ alone starts to dominate the
RGEs~(\ref{eq:HSM}),(\ref{eq:HMSSM}).
The precise form of the $M_\nu$ dependence
is quite sensitive to the ``angles'' $y_{ij}$, but the roughly 
linear growth of $|\sum_i \epsilon_i|$ in the regime of
large $M_\nu$ appears to be general. However, the figure also clearly
shows a strong dependence on the MSSM parameter $\tan\beta$
particularly for small $M_\nu$. Indeed
already for relatively small $\tan\beta=10$ the CP asymmetries can be
more than an order of magnitude larger than in the SM. Moreover, in the
case of the MSSM we observe a sign change at some scale
$M_\nu \stackrel{>}{\sim}\, 10^{12}$ GeV,
which can be traced to the different relative signs between the terms
on the right-hand sides of (\ref{eq:HSM}) and (\ref{eq:HMSSM}).
This example clearly demonstrates a rather dramatic dependence on
details of the model.
Finally, as in the case of the double ratios above, we investigated
the impact of the iterative procedure compared to the simplified
approach and found it to be generically small. Hence we feel justified
to use the simplified procedure in Section~\ref{sec:leptogenesis} in
order to save computer time.

\section{Leptogenesis in the extended MLFV Framework}
\label{sec:leptogenesis}
\subsection{Preliminaries}

One of the most plausible mechanisms for creating the observed
matter--antimatter asymmetry in the universe is leptogenesis, where a CP
asymmetry generated through the out-of-equilibrium $L$-violating decays of the
heavy Majorana neutrinos leads to a lepton asymmetry which is subsequently
transformed into a baryon asymmetry by $(B+L)$-violating sphaleron processes
\cite{Fukugita:1986hr,Buchmuller:2005eh,Hambye:2004fn}.

Unfortunately,
even in its simplest realization through the well-known seesaw
mechanism~\cite{Minkowski:1977sc}, the theory has too many
parameters. Indeed, as recalled in Section 2.4  in the framework of
the standard model (SM) extended with three heavy Majorana neutrinos $N_{i}\,
(i=1,2,3)$, the high-energy neutrino sector, characterized by the Dirac
neutrino ($m_D$) and the heavy Majorana neutrino ($M_R$) mass matrices, has
eighteen parameters. Of these, only nine combinations enter into the seesaw
effective neutrino mass matrix $m_D^T\,M_R^{-1}\,m_D\,$, thus making difficult
to establish a direct link between leptogenesis and low-energy
phenomenology~\cite{Branco:2001pq}. Furthermore, there are six CP-violating
phases which are physically relevant at high energies, while only three
combinations of them are potentially observable at low energies. Therefore, no
direct link between the sign of the baryon asymmetry and low-energy leptonic
CP violation can be established, unless extra assumptions are introduced.

Furthermore, additional
assumptions are usually required to completely determine the high-energy
neutrino sector from low-energy observables. Typical examples are the
introduction of texture zeros in the Yukawa matrices or the imposition of
symmetries to constrain their structure~\cite{Kaneko:2002yp}. On the other
hand,
the heavy Majorana neutrino masses can range
from the TeV region to the GUT scale, and the spectrum can be hierarchical,
quasi-degenerate or even exactly degenerate~\cite{GonzalezFelipe:2001kr}.
Despite this arbitrariness, the heavy Majorana neutrino mass scale 
turns out to be crucial for a successful
implementation of the leptogenesis mechanism. In particular, the standard
thermal leptogenesis scenario with hierarchical heavy Majorana neutrino masses
($M_1 \ll M_2 < M_3$) requires $M_1 \gtrsim 4 \times
10^8$~GeV~\cite{Hamaguchi:2001gw}, if $N_{1}$ is in thermal equilibrium before
it decays, or the more restrictive lower bound $M_1 \gtrsim 2 \times
10^9$~GeV~\cite{Giudice:2003jh} for a zero initial $N_{1}$ abundance. Since
this bound also determines the lowest reheating temperature allowed after
inflation, it could be problematic in supersymmetric theories due to the
overproduction of light particles like the gravitino~\cite{GravitinoProblem}.

It should be emphasized, that the above bounds are  model
dependent in the sense that they can be avoided, if the heavy Majorana
neutrino spectrum is no longer hierarchical. For example, if at least two of the
$N_{i}$ are quasi-degenerate in mass, \emph{i.e.} $M_1 \simeq M_2\,$, then the
leptonic CP asymmetry relevant for leptogenesis exhibits the resonant
behavior $\varepsilon_1 \sim
M_1/(M_2-M_1)$~\cite{Pilaftsis:1997jf,Pilaftsis:2003gt}. In this case, it is
possible to show that the upper bound on the CP asymmetry is independent of
the light neutrino masses and successful leptogenesis simply requires $M_{1,2}$
to be above the electroweak scale for the sphaleron interactions to be
effective. The quasi-degeneracy may
also be achieved in soft leptogenesis where a small splitting is induced
by the soft supersymmetry breaking terms~\cite{D'Ambrosio:2003wy}.

Another possibility which has been recently
explored~\cite{GonzalezFelipe:2003fi,Turzynski:2004xy} relies on the fact that
radiative effects, induced by the renormalization group (RG) running from high
to low energies, can naturally lead to a sufficiently small neutrino mass
splitting at the leptogenesis scale. In the latter case, 
sufficiently large CP asymmetries are generated.

In the minimal seesaw scenario with only two heavy
neutrinos the
resulting baryon asymmetry in the SM turns out to be below the
observed value~\cite{GonzalezFelipe:2003fi}. On the other hand, this mechanism can be successfully
implemented in its minimal supersymmetric extension
(MSSM)~\cite{Turzynski:2004xy}. 

It has been shown \cite{Branco05} that the above problems in the SM can be overcome in a more 
realistic scenario where the effects of a third heavy neutrino are also taken
into account. In \cite{Branco05}, 
leptogenesis was studied in the framework of a model
where there are three right-handed neutrinos, with masses 
$M_1\approx M_2\ll M_3$. We will discuss this scenario below as a special
limit of the MLFV framework.

In view of the above, it is important to analyze leptogenesis in the extended
MLFV framework, where CP violation is allowed both at high and low energies. 
In the MLFV scenario, right-handed neutrinos are assumed to be {\it exactly}
degenerate at a high energy scale. In the limit of exact degeneracy, no
lepton-asymmetries can be generated. However, as previously emphasized, even
if exact degeneracy is assumed at a high energy scale, renormalization group
effects lead to a splitting of right-handed neutrino masses at the scale of
leptogenesis, thus offering the possibility of viable leptogenesis in the
extended MLFV framework.

\subsection{BAU in the RRL and Flavour Effects}
In leptogenesis scenarios the baryon asymmetry of the universe $\eta_B$ 
arises due to non-perturbative sphaleron interactions that turn a lepton asymmetry into a
baryon asymmetry. The predicted value of $\eta_B$ has to match the results of 
WMAP and the BBN analysis for the primordial deuterium abundance \cite{Spergel:2003cb}

\be
\eta_B=\frac{n_B}{n_\gamma}=(6.3\pm 0.3)\times 10^{-10}.
\ee

The lepton asymmetry is generated by
out-of-equilibrium decays of heavy right-handed Majorana neutrinos
$N_i$ and is proportional to the CP asymmetry $\varepsilon_i^l$ with
\be        \label{eq:cpas}
\varepsilon_i^l=\frac{\Gamma(N_i\to L_l\,\phi)-\Gamma(N_i\to \bar L_l\, \bar \phi)}{\sum_l\left[\Gamma(N_i\to L_l\,\phi)+\Gamma(N_i\to \bar L_l \,\bar \phi)\right]}\,,
\ee
and $l$ denoting the lepton flavour, that arises at one-loop order due to the
interference of the tree level amplitude with vertex and self-energy
corrections.\\
A characteristic of the MLFV framework is that 
only admissible BAU with the help of
leptogenesis
is radiative and thereby resonant leptogenesis. The mass
splittings of the right-handed neutrinos induced by the RGE
are of similar size $\Delta M \sim \mathcal{O}(M_\nu \,Y_\nu
Y_\nu^\dagger)$ as the decay widths $\Gamma \sim  \mathcal{O}(M_\nu \,Y_\nu
Y_\nu^\dagger)$. This is the 
condition of resonant leptogenesis. 
The CP asymmetry is for the lepton flavour $l$ given by

\be
	\varepsilon_i^l = \frac{1}{(Y_\nu Y_\nu^\dagger)_{ii}} \sum_j \Im ( (Y_\nu Y_\nu^\dagger)_{ij}   (Y_\nu )_{il} (Y_\nu^\dagger)_{lj} ) \, g( M_i^2,M_j^2 ,\Gamma_j^2 )
\ee
where $g( M_i^2,M_j^2 ,\Gamma_j^2 )$ is an abbreviation for the full result
given in \cite{Pilaftsis:2003gt}. The total CP asymmetries $\varepsilon_i$ are obtained
summing over the lepton flavours $l$.\\
The baryon to photon number ratio $\eta_B$  can then be calculated solving the
Boltzmann equations for the lepton asymmetry and converting it into
$\eta_B$ using suitable dilution and sphaleron conversion
factors. Which Boltzmann equation to use depends on the temperature
scale at which leptogenesis takes place. We will follow a simplistic
approach ignoring all subtleties generically coming into play in the
intermediate regime between different mechanisms at work. Our main
conclusions, however, will not be affected by this omission. We will
simply divide the temperature scale into a region up to which all
three lepton flavours have to be taken into account and a region above
which the single flavour approximation works.

Below some temperature\footnote{
We will chose $T^\mu_{\rm eq}\simeq 10^{10}$ GeV in our analysis as
an effective boundary between the unflavoured and `fully flavoured'
regimes, where we (respectively) neglect flavour and distinguish all
three flavours. The main conclusions are, however, not affected by the
precise choice. }
$T^\mu_{\rm eq}  \simeq 10^{9}$
  GeV~\cite{Barbieri:1999ma,Abada:2006fw,Nardi:2006fx}, muon and tau charged lepton
Yukawa interactions are much faster than the expansion $H$ rendering
the $\mu$ and $\tau$ Yukawa couplings in equilibrium. The correct
treatment in this regime requires the solution of lepton
flavour specific
Boltzmann equations. 
In the strong washout regime $\eta_B$ is independent of the initial abundances and 
an estimate including flavour effects
is given by~\cite{Pilaftsis:2005rv} 
\be\label{etaB-estimate}
\eta_B\simeq -10^{-2} \,\sum_{i=1}^3\sum_{l=e,\mu,\tau}\,e^{-(M_i-M_1)/M_1}\,
\varepsilon_i^l\,\frac{K_i^l}{K^l K_i},
\ee
with 
\be
K_i^l=\frac{\Gamma(N_i\to L_l \phi)+\Gamma(N_i \to \bar L_l\bar\phi)}{H(T=M_i)}
\ee
\be
K_i=\sum_{l=e,\mu,\tau}K_i^l,\qquad K^l=\sum_{i=1}^3 K_i^l, \qquad
H(T=M_i)\simeq 17 \frac{M_i^2}{M_{\rm Pl}} 
\ee
where $M_{\rm Pl}=1.22\times 10^{19}$ GeV and $K_i^l$ is the washout factor due to the inverse decay of the Majorana
neutrino $N_i$ into the lepton flavour $l$. 
The impact of lepton flavour effects on $\eta_B$ is discussed in 
 \cite{Barbieri:1999ma,Pilaftsis:2005rv,Abada:2006fw,Nardi:2006fx,Blanchet:2006be}. As we shall also see below, the inclusion of flavour
effects generally leads to an enhancement of the resulting
$\eta_B$. This is due to two effects: (1) the washout gets reduced
since the interaction with the Higgs is with the flavour
eigenstates only and (2) an additional source of CP violation arises
due to lepton flavour specific CP asymmetries.
  
For higher values of $T \stackrel{>}{{}_\sim} 10^{12}$ GeV the charged
lepton Yukawa couplings do not break the coherent evolution of the
lepton doublets produced in heavy neutrino decays anymore. In this regime
flavour effects can be ignored and an order of magnitude estimate is given
by
\be\label{etaB-estimate_nf}
\eta_B\simeq -10^{-2} \,\sum_{i=1}^3\,e^{-(M_i-M_1)/M_1}\,\frac{1}{K} \,\sum_{l=e,\mu,\tau}
\varepsilon_i^l,
\ee
with $K = \sum_i K_i$. 
This agrees with a recent analytical estimate
by~\cite{diBari} up to factors of $\mathcal{O}(1)$ for the region of
interest in parameter space, where the estimate of~\cite{diBari}
generally leads
to a smaller efficiency and smaller $\eta_B$. We have also compared the analytical estimate
of~\cite{diBari} and~(\ref{etaB-estimate_nf}) with the numerical
solution of the Boltzmann equations using the
LeptoGen
code~\cite{Pilaftsis:2005rv}\footnote{http://www.ippp.dur.ac.uk/$\sim$teju/leptogen/}. For
the relevant ranges of the input parameters the analytical estimate of
~(\ref{etaB-estimate_nf}) and the full numerical solution agree quite
well, whereas the estimate of~\cite{diBari} leads to an efficiency and $\eta_B$
generally smaller by a factor of 5 to 10. This is shown in Fig.~\ref{fig:3flavor_MR_comp}.
 These estimates, however, do not take
into account the potentially large lepton flavour effects included
in~(\ref{etaB-estimate}).\\
Let us remark in passing that in the flavour independent region we are always in the strong
washout regime, since
\be\label{strong_washout_nf}
K = K_1 + K_2 + K_3 = \frac{1}{m^*}\, {\rm tr } \left( R d_\nu R^\dagger\right) 
\ge \frac{ (\Delta m_{\rm atm}^2)^{1/2}}{m^*} \simeq 50,
\ee
where $m^* = \mathcal{O}(10^{-3})$. This inequality holds since the
trace is linear function of the neutrino masses with positive
coefficients, which reaches its minimum for $y_{ij}=0$. We also made sure
that the estimate~(\ref{etaB-estimate})
including flavour effects is applicable~\cite{Pilaftsis:2005rv}
and checked that the inequality
\be\label{strong_washout}
K^l_i  \stackrel{>}{{}_\sim} 1
\ee
is always satisfied for the points considered in the
plots. Since both~(\ref{strong_washout_nf}) and~(\ref{strong_washout}) are satisfied, a
simple decay-plus-inverse decay picture is a good description and
the estimates~(\ref{etaB-estimate}) and~(\ref{etaB-estimate_nf})
independent of the initial abundances
give a good approximation of the numerical solution of the full Boltzmann
equations. \\
We have performed the leptogenesis analysis specifically for the
SM. We do not
expect large deviations in the MSSM from the SM if the same $Y_\nu(M_\nu)$ and $M_\nu^i(M_\nu)$ are given. The main differences
come (1) from the CP-asymmetries, which now include contributions
from the supersymmetric particles, (2) from the washout,
and (3) from conversion and dilution factors. 
The supersymmetric CP asymmetries have the same
flavour structure as in the SM and using~\cite{Covi:1996wh} one can
show that $\epsilon^{MSSM}\simeq 2\,
\epsilon^{SM}$ for quasi-degenerate heavy neutrinos. We also expect the
correction by the decay widths to be similar in size. 
Next, the washout in the strong washout regime is about a factor of
$\sqrt{2}$ larger~\cite{DiBari:2004en} in the MSSM, whereas the dilution and
sphaleron conversion factors stay almost unchanged. Concluding, we find that in the scenario considered the
predicted values roughly satisfy $\eta_B^{\rm MSSM} \simeq
1.5 \,\eta_B^{\rm SM}$ for the same set of
input parameters $Y_\nu(M_\nu)$ and $M_\nu^i(M_\nu)$. The RGE induced values of $Y_\nu(M_\nu)$ and $M_\nu^i(M_\nu)$, however, are model dependent and lead to in general different $Y_\nu(M_\nu)$ and $M_\nu^i(M_\nu)$ for the same boundary conditions at the GUT and low-energy scale, as discussed in Section~\ref{sec:RGE-high}. Especially sensitive is the region $M_\nu  \stackrel{<}{{}_\sim}
10^{12}$ GeV where the CP asymmetries are dominantly generated by the tau Yukawa coupling, which is enhanced by a factor of $\tan \beta$ in the MSSM. 
Note also that in the MSSM,
$T^\mu_{\rm eq}$ and $T^\tau_{\rm eq}$ should be rescaled by a factor $(1 + \tan^2\beta)$ to
take account of the larger Yukawa couplings~\cite{Antusch:2006cw}, which should
make flavour effects even more prominent.\footnote{We thank S.~Antusch for
drawing our attention to this point.}

\subsection{Two flavour limit}

\vspace{0.2cm}

\begin{figure}[h]
\vspace{0.01in}
\centerline{
\epsfysize=3in
\epsffile{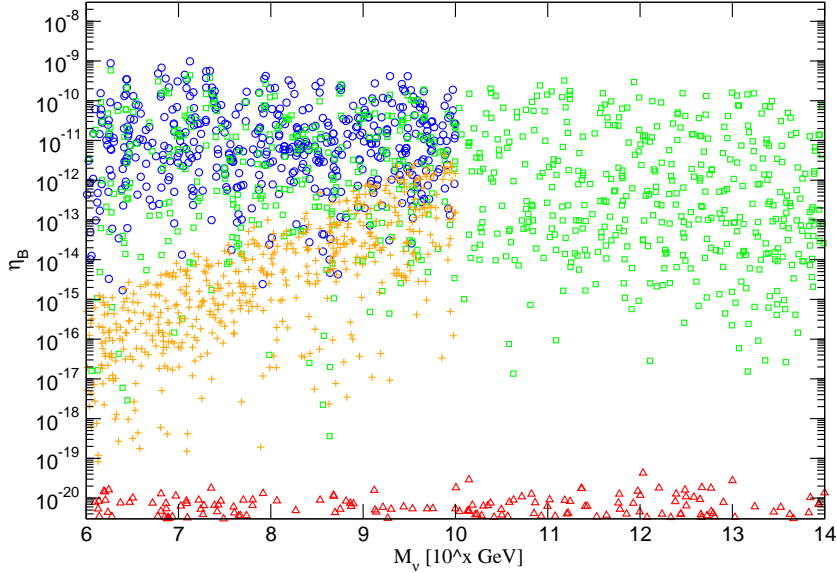}}
\caption{Resulting $\eta_B$ for the case in which only $y_{12}\not= 0$
  (effective two flavour case) as a function of $M_{\nu}$ for the
  normal hierarchy of light neutrinos: the \emph{orange crosses} and \emph{red triangles} show the unphysical limit setting the charged lepton Yukawas $Y_e = 0$ in the renormalization group evolution with and without including lepton flavour effects in the calculation of $\eta_B$, respectively. Setting the charged lepton Yukawas to their physical values, the \emph{blue circles} and the \emph{green squares} correspond to including and ignoring lepton flavour effects in the calculation, respectively.}\label{fig:2flavMR}
\end{figure}

\begin{figure}[h]
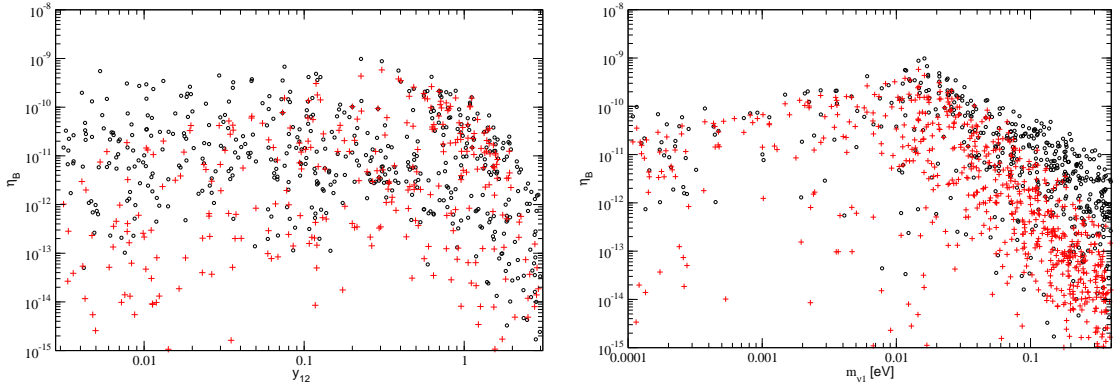

\centerline{
\epsfysize=2in
\epsffile{2flavor_y12_selected.eps}
\hspace{0.07in}
\epsfysize=2in
\epsffile{2flavor_mv_selected.eps}}
\caption{$\eta_B$ for the case in which only $y_{12}\not= 0$
  (effective two flavour case) as a function of $y_{12}$ (left) and
  $m_{\nu 1}$ (right) for the normal hierarchy. The \emph{black
    circles} are obtained including lepton flavour effects and the
  \emph{red crosses} are calculated ignoring them.}\label{fig:2flavy12}
\end{figure}

As a first step we discuss the special case of $y_{12}$ being
non-vanishing at the GUT scale and all other $y_{ij}=0$. This
corresponds approximately to one of the scenarios considered in a
recent study of radiative leptogenesis~\cite{Branco05} with two
right-handed neutrinos quasi-degenerate and a third right-handed
neutrino decoupled $M_1 \simeq M_2\, \ll M_3$. If only $y_{12} \not=
0$ the calculation of $\eta_B$ proceeds in the same way since
to a good approximation only $\nu^1_R$ and $\nu^2_R$ contribute to the
CP asymmetry. The only difference comes from the enhanced
wash-out. Since the third heavy neutrino is now also contributing, the lower
bound on the washout $K$ in ~(\ref{strong_washout_nf}) is in our case relatively
enhanced by a factor $ (\Delta m_{\rm
  atm}^2)^{1/2}/ (\Delta m_{\rm
  sol}^2)^{1/2} \simeq 4-5$. We have checked this correspondence for $\eta_B$ also
numerically.
Ignoring flavour subtleties in leptogenesis for a moment, the CP violating effects due to renormalization group effects are induced only by the charged lepton yukawa couplings, see Section~\ref{sec:RGE-high}, and the total CP asymmetries for each heavy Majorana neutrino take the form~\cite{Branco05}
\begin{align}
\label{epsigen}%
    \varepsilon_{1,2} \simeq \frac{ \bar\varepsilon_{1,2}}{1 + D_{2,1}}, \qquad
     \varepsilon_{3} \simeq 0,
\end{align}
and 
\begin{align}
    \label{CPAsymmetry_uncorrected}
    \bar\varepsilon_j &\simeq \frac{3y^2_\tau}{32\, \pi}
    \frac{\text{Im} (H_{21})\, 
    \text{Re} \left[(Y_\nu)^\ast_{23}\, (Y_\nu)_{13}\right]}
     {H_{jj}(H_{22}-H_{11})}\, = \frac{3\, y^2_\tau}{64\, \pi}\,
  \frac{m_j (m_1 + m_2)\, \sqrt{m_1\, m_2}\, \sinh(2\, y_{12})\,
  \text{Re}\, (U^\ast_{\tau 2}\, U_{\tau 1})}{ (m_1 - m_2)(m_{j }^2\, \cosh^2 y_{12} +
  m_{1} m_{2}\, \sinh^2 y_{12})}\, , \\
    D_j &\simeq \frac{\pi^2}{4} \frac{H^2_{jj}}{(H_{22} - H_{11})^2\, \ln^2 \left(M_\nu/M_{\rm GUT}\right)} \,= \left[\frac{\pi}{2}\, \frac{m_{j}^2\, \cosh^2 y_{12} + m_{2} m_1\, \sinh^2 y_{12}}
	  {m_j (m_2 - m_1)\, \ln \left(M_\nu/M_{\rm GUT}\right)}
	  \right]^2\,  .
    \label{Dj}
\end{align}

where $D_j$ are regularization factors coming from the heavy Majorana decay widths. We immediately see that the total CP asymmetries only bare a very mild dependence on the heavy Majorana scale. The almost negligible dependence on $M_\nu$ has to be compared with the power-suppression in $M_\nu$ in the hierarchical case ($M_1 \ll M_2 < M_3$). We find this expectation confirmed in Fig. \ref{fig:2flavMR}, where the resulting $\eta_B$ is shown as a function of $M_\nu$.\\
Fig. \ref{fig:2flavMR} also nicely illustrates the
relative importance of flavour effects in leptogenesis. If no cancellations occur, we find, that flavour effects
generate an $\eta_B$ which is of the same order of
magnitude (blue circles), however almost always larger than the one calculated
ignoring flavour effects (green squares). \\
If we now consider the unphysical limit of setting $Y_e = 0$ in the renormalization group running only, we find that the \emph{total} CP asymmetries and $\eta_B$ should vanish since no CP violation effects are induced by the RGE, see Section~\ref{sec:RGE-high}. We confirm this behavior in Fig. \ref{fig:2flavMR} (red triangles). 
A very different picture emerges once we include flavour effects. The
relevant quantity for leptogenesis is then $ \Im ( (Y_\nu
Y_\nu^\dagger)_{ij}   (Y_\nu )_{il} (Y_\nu^\dagger)_{lj} ) $ with no
summation over the charged lepton index $l$. Although no \emph{total}
CP asymmetries are generated via the RG evolution in the limit $Y_e =
0$, the CP asymmetries for a specific lepton flavour are
non-vanishing. 
Additionally, the resulting $\eta_B$ now shows a $M_\nu$
dependence which stems from the RGE contributions due to $Y_\nu$
only, which are absent in the total CP asymmetries in the two flavour
limit (orange crosses). \\
All plots have been generated assuming a normal hierarchy
of the light neutrino masses. We have checked that the results for the
inverted hierarchy are similar, although $\eta_B$ turns out to be generally 
smaller and below the observed value, in accordance with the findings
of~\cite{Branco05}. Including flavour effects it is however still
possible to generate a $\eta_B$ of the correct order of magnitude.
 In
Fig.~\ref{fig:2flavy12} we additionally show the dependence of $\eta_B$ on $y_{12}$ and
$m_{\nu 1}$. We find that flavour effects enlarge the $y_{12}$ range
where successful baryogenesis is possible and slightly soften the upper bound
on the light neutrino mass scale. The left panel even demonstrates that leptogenesis in the MLFV scenario
is possible for a real $R$ matrix. Then lepton flavour effects are
essential for a successful leptogenesis~\cite{Abada:2006fw,Nardi:2006fx}.

\subsection{General case}
\begin{figure}[h]
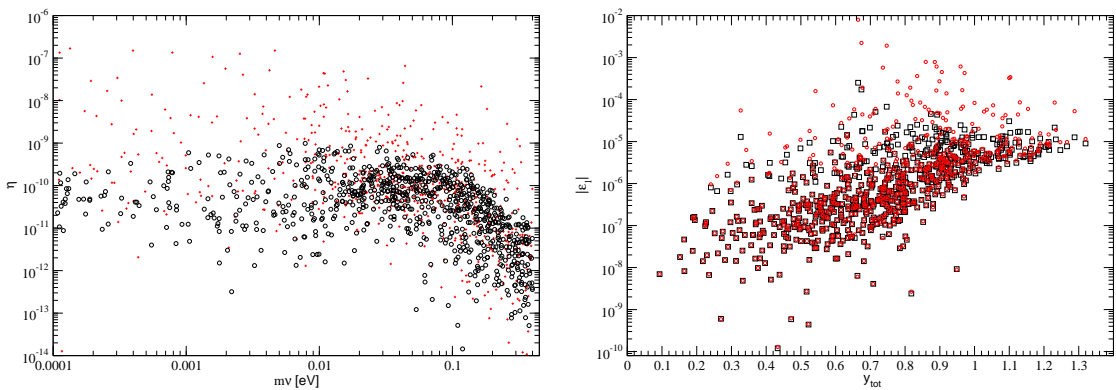

\vspace{0.1in}
\centerline{
\epsfysize=2in
\epsffile{3flavor_mv_selected.eps}
\hspace{0.07in}
\epsfysize=2in
\epsffile{cpa_ytot_3flavor.eps}}
\caption{(left) $\eta_B$ for the general case with $0.01 < \left| y_{ij} \right| < 1$
  as a function of $m_{\nu 1}$ (right). The \emph{black circles} are
  obtained including lepton flavour effects whereas the \emph{red crosses}
  are calculated ignoring them. The flavour blind results (\emph{red
    crosses}) reach higher values due to the CP asymmetries growing as $M_\nu$
  gets bigger in this regime.
 (right) The total CP asymmetry
  $|\epsilon_1|$ for the general case with $0.01 < \left| y_{ij}\right|  < 0.8 $ as a
  function of $y_{tot} = (y_{12}^2 + y_{13}^2 + y_{23}^2)^{\frac12}$
  for input values that result in the right oder of magnitude of
  $\eta_B$. The \emph{red circles} are obtained using the uncorrected
  CP asymmetries and the \emph{black squares} include the corrections
  by the decay widths}\label{fig:3flav}
\end{figure}

Now we consider the general case with all three phases $y_{ij}$
non-vanishing. We have varied the parameters as described in
Section \ref{sec:numerics}. 
The regularization of the resonant CP asymmetry by the $D_i$ turns out
to be important for values of $\epsilon_i \stackrel{>}{{}_\sim}
10^{-6}$, see Fig.~\ref{fig:3flav}. As seen there, in the regime where flavour effects
are important we find an upper bound on the light neutrino mass of
$m_{\nu 1}\stackrel{<}{{}_\sim} 0.2$ eV in order to generate the right
amount of $\eta_B$. Beyond the temperature scale where flavour
effects play a role, no relevant bound can be found. This is due to
the enhancement of the CP asymmetry which approximately grow linearly for values of $M_\nu  \stackrel{>}{{}_\sim}
10^{12}$ GeV, see the discussion in Section~\ref{sec:CPA-numerics} \\

\vspace{0.1cm}
\begin{figure}[h]
\vspace{0.01in}
\centerline{
\epsfysize=3in
\epsffile{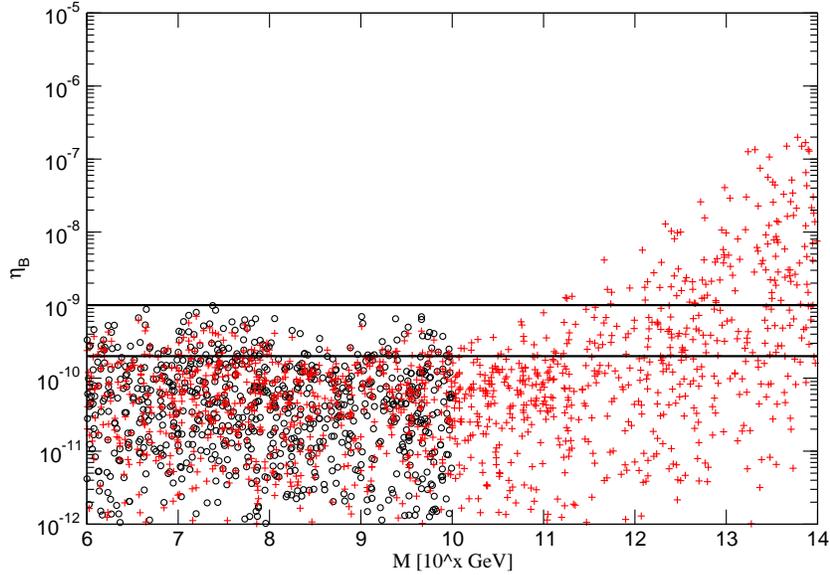}}
\caption{$\eta_B$ for the general case with $0.01 < \left| y_{ij}\right| < 1 $ as a function of $M_{\nu}$. The \emph{circles} are obtained including lepton flavour effects and the \emph{red crosses} are calculated ignoring them. }\label{fig:etaB}
\end{figure}

\begin{figure}[ht]

\centerline{
\epsfysize=3in
\epsffile{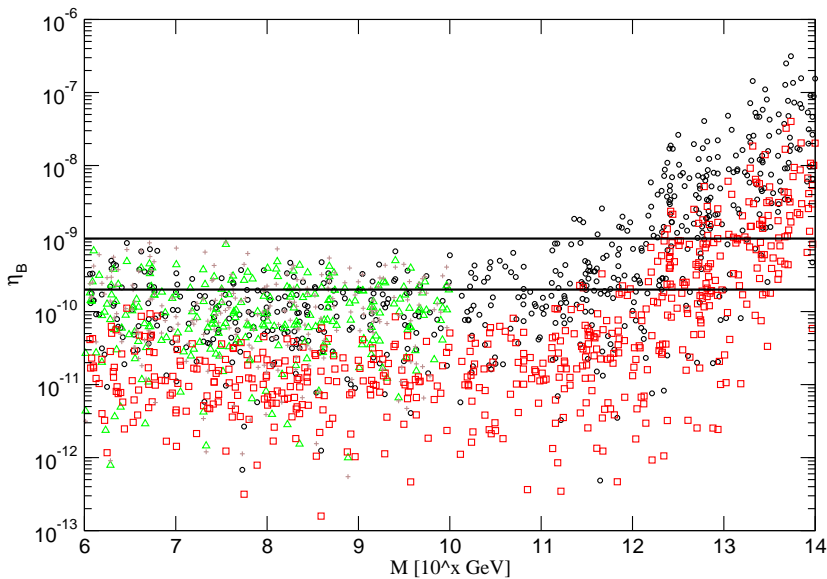}
}
\caption{Different determinations of $\eta_B$ for the general case with $0.01 < \left| y_{ij}\right|  < 1$ as a
  function of $M_{\nu}$. The \emph{black circles} are obtained
  numerically solving the flavour independent Boltzmann equations using the
  LeptoGen package, the \emph{green triangles} and the \emph{brown
    crosses} show estimates including flavour effects of~\cite{Blanchet:2006be} and
  ~(\ref{etaB-estimate}), respectively. Finally the \emph{red boxes} show the estimate of~\cite{diBari}
   used in Cirigliano et. al~\cite{CIP06} which ignores flavour effects.}\label{fig:3flavor_MR_comp}
\end{figure}

\vspace{0.1cm}
In Fig.~\ref{fig:3flavor_MR_comp}, we compare different calculations
of $\eta_B$:
\begin{itemize}
\item
the flavour independent estimate of \cite{diBari} used in 
Cirigliano et al.~\cite{CIP06} (red boxes),
\item
the numerical solution of the flavour independent Boltzmann equations 
using the LeptoGen package (black circles),
\item
the recent estimate by Blanchet and Di Bari \cite{Blanchet:2006be} that
includes flavour effects (green triangles),
\item
the approximate expression of \cite{Pilaftsis:2005rv} given in (\ref{etaB-estimate}) that
also includes flavour effects (brown crosses).
\end{itemize} 

We find that
\begin{itemize}
\item
the flavour blind estimate of \cite{diBari} used in Cirigliano
et. al.~\cite{CIP06} lies consistently below
the numerical solution of the flavour independent Boltzmann
equations. For $M_\nu\ge 10^{12}\gev$ this turns 
out to be unimportant as flavour effects in this region are small and 
we confirm the increase of $\eta_B$ with $M_\nu$ in this region 
found by these authors.
\item
Potentially large
flavour effects that have been left out in \cite{CIP06} generally
enhance the predicted $\eta_B$, in particular for $M_\nu\le 10^{12}\gev$, in accordance with the existing 
literature.
\item
Both flavour estimates and the numerical
solution of the flavour independent Boltzmann equations show solutions 
with $\eta_B$ of the
measured order of magnitude without imposing a stringent lower bound on the value of
$M_{\nu}$. 
\end{itemize}

The last finding is in contrast to the analysis of Cirigliano
et. al.~\cite{CIP06} which using the flavour independent estimate of 
\cite{diBari} finds a lower bound on $M_{\nu}$ of $\mathcal{O}(10^{12}{\rm GeV})$ as clearly represented 
by the red boxes  in
Fig.~\ref{fig:3flavor_MR_comp}. The same qualitative conclusion holds for $\eta_B$ using the RGE induced CP asymmetries in the MSSM. The $\tan \beta$ enhancement of the CP asymmetries as discussed in Section~\ref{sec:CPA-numerics} even facilitates the generation of an $\eta_B$ of the right size.

Our analysis that includes flavour effects demonstrates that
baryogenesis through leptogenesis in
the framework of MLFV is a stable mechanism and allows a successful 
 generation of $\eta_B$
over a wide range of parameters. The absence of a lower bound on $M_{\nu}$ 
found here has of course an impact on the LFV processes, which we will discuss
next.

\vspace{1 truecm}

\vspace{0.2cm}
\begin{figure}[h]
\centerline{
\epsfysize=3.5in
\epsffile{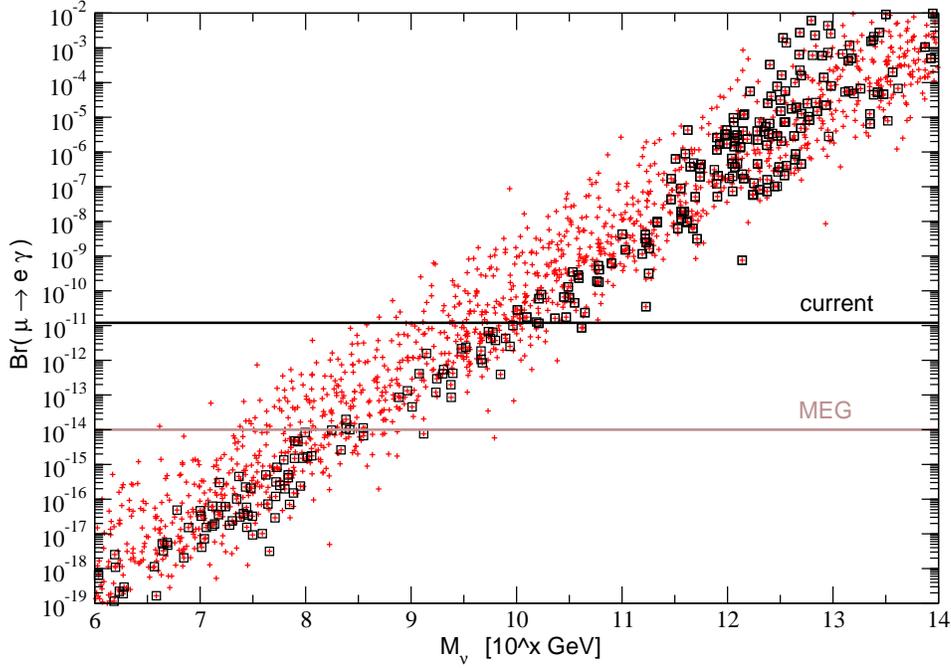}}
\caption{$B(\mu \rightarrow e \gamma)$ as a function of $M_\nu$ for
  $\Lambda_{\rm LFV} =1$ TeV. The \emph{black squares} show points
  where a baryon asymmetry in the range $2 \cdot 10^{-10} < \eta_B <
  10\cdot 10^{-10} $ is possible. }\label{fig:muegamma}
\end{figure}

In Fig.~\ref{fig:muegamma} we
show $B(\mu \rightarrow e \gamma)$ vs. $M_\nu$ for the parameter ranges described above and a lepton flavor violation scale of $1$ TeV. We highlighted the points where 
successful baryogenesis is possible (black squares). We find that $B(\mu \rightarrow e \gamma)$ can be made small enough to evade bounds from current and future experiments and one can have successful baryogenesis through leptogenesis at the same time. This is another finding of our paper which is in contrast to a recent analysis \cite{CIP06}. We will summarize the differences to \cite{CIP06} in the next paragraph.

\subsection{Comparison with \cite{CIP06}}
\label{sec:comparison}
Recently in an independent analysis Cirigliano, Isidori and 
Porretti \cite{CIP06} generalized MLFV formulation in \cite{MLFV} to include 
CP violation at low and high energy. Similarly to us they found it convenient
to use for $Y_\nu$ the parametrization of Casas and Ibarra. They
have also pointed out that in the MFLV framework the most natural is the 
resonant leptogenesis.

On the other hand, these authors
neglected flavour dependent effects in the evaluation of $\eta_B$, 
that we find in agreement with other authors to be important~\cite{Barbieri:1999ma,Pilaftsis:2005rv,Abada:2006fw,Nardi:2006fx,Blanchet:2006be}. This has important consequences already at the qualitative
level.
Their qualitative discussion of the splittings of the $M_\nu^i$ at the 
see-saw scale is similar to ours and we agree with the main physical points made by these authors in this context.
On the other hand, while we have demonstrated 
explicitely by means of a renormalization group analysis that a 
successful RRL can be achieved, Cirigliano et al confined their analysis 
to parametrizing possible radiative effects in terms of a few parameters.
In this context a new point made by us (see discussion Section~\ref{sec:largelog}) is that the coefficients $c_i$ in (\ref{eq:onshell}) are in fact not independent of each other. Indeed the leading logarithmic contribution to $c_i$ are related by the renormalization group. This can in principle increase the predictivity of MLFV.\\
The three most interesting messages of \cite{CIP06} are
\begin{itemize}
\item
A successful resonant leptogenesis within the MLFV framework implies a lower
bound $M_\nu\ge 10^{12}\gev$,
\item
With $\Lambda_{\rm LFV}=\ord(1\tev)$, this lower bound implies the rate for 
$\mu\to e\gamma$ close to the present exclusion limit,
\item
MLFV implies a specific pattern of charged LFV rates: 
$B(\mu\to e \gamma) <B(\tau\to\mu\gamma)$.
\end{itemize} 
For $M_\nu\ge 10^{12}\gev$, in spite of some differences in the numerics 
 as discussed above, we basically confirm these findings. Unfortunately, 
 for lower values of $M_\nu$ our results differ from theirs. In particular, 
as we have demonstrated in Fig.~\ref{fig:3flavor_MR_comp}, the observed value of $\eta_B$ can be 
obtained for $M_\nu$ by several orders of magnitude below the bound in 
\cite{CIP06}, in accordance with other analyses of leptogenesis. Once 
$M_\nu$ is allowed to be far below $10^{12}\gev$, $\Lambda_{\rm LFV}=\ord(1\tev)$
 does not 
imply necessarily $B(\mu\to e\gamma)$ close to the inclusion limit 
as clearly seen in Fig.~\ref{fig:muegamma}. 

One of the reasons for the discrepancy between our result with regard 
to $M_\nu$ and the one of \cite{CIP06} is the neglect of flavour effects in 
leptogenesis in the latter paper. Fig.~\ref{fig:3flavor_MR_comp} illustrates that flavour 
effects in leptogenesis matter.

Concerning $B(\mu\to e \gamma) <B(\tau\to\mu\gamma)$, we confirm the 
result of \cite{CIP06} in the limit of very small $y_{12}$, but as
shown in Fig.~\ref{fig:LFVdoubleratio}, this is not true in general, as also found in~\cite{Kanemura:2005cq,Petcov:2006bn}.
Consequently, this hierarchy of charged LFV rates cannot be used as 
model independent signature of MLFV.

\section{Summary and Conclusions}
\label{sec:conclusions}
In this paper we have generalized the proposal of minimal
flavour violation in the lepton sector of \cite{MLFV} to include CP
Violation at low and high energy. While the definition
proposed in \cite{MLFV} could be considered to be truly minimal, it appears
to us too restrictive and not as general as the one in the
quark sector (MFV) in which CP violation at low energy is
automatically included \cite{UUT} and in fact all flavour violating
effects proceeding through SM Yukawa couplings are taken
into account \cite{AMGIISST}. The new aspect of MLFV in the
presence of right-handed neutrinos, when compared with MFV,
is that the driving source of flavour violation, the neutrino Yukawa
matrix $Y_\nu$, depends generally also on physics at very high
scales. This means also on CP violating sources relevant for
the generation of baryon-antibaryon asymmetry with the help
of leptogenesis. The first discussion of CP violation at low and high energy has been
 presented in \cite{CIP06}. Our conclusions for $M_\nu \ge  10^{12}$ agree basically
 with these authors. However, they differ in an essential manner for lower
 values of $M_\nu$.

The main points of our paper have been already summarized in
the introduction. Therefore it suffices to conclude our
paper with the following messages:
\begin{itemize}
\item
A new aspect of our paper is the realization that in
the context of MLFV the only admissible BAU with the help of
leptogenesis is the one through radiative resonant leptogenesis (RRL). Similar observations have been made in~\cite{CIP06}.
In this context our analysis benefited from the ones in
 \cite{GonzalezFelipe:2003fi,Turzynski:2004xy,Pilaftsis:1997jf,Pilaftsis:2003gt}.
The numerous analyses of leptogenesis with hierarchical
right-handed neutrinos present in the literature are
therefore outside the MLFV framework and the
differences between the results presented here and the ones
found in the literature for $M_1\ll M_2\ll M_3$ can be used to distinguish
MLFV from these analyses that could be affected by new flavour
violating interactions responsible for hierarchical right-handed neutrinos.
\item
We have demonstrated explicitely within the SM and the MSSM at low $\tan\beta$ 
that within a general MLFV scenario the right size of $\eta_B$ can indeed 
be obtained
by means of RRL.
The important property of this type of leptogenesis is the 
very weak sensitivity of $\eta_B$ to the see-saw scale $M_\nu$ so that for 
scales as low as $10^6\gev$ but also as high as $10^{13}\gev$, 
the observed $\eta_B$ can be found.
\item
 Flavour effects, as addressed by several authors recently
in the
literature~\cite{Barbieri:1999ma,Pilaftsis:2005rv,Abada:2006fw,Nardi:2006fx,Blanchet:2006be},
play an important role for $M_\nu \sle 10^{10}$ GeV as they generally
enhance $\eta_B$. Moreover, they allow for a
successful leptogenesis within MLFV even when the $R$-matrix is real
(left panel of Fig.~\ref{fig:2flavy12}).
\item
As charged LFV processes, like $\mu\to e\gamma$ are sensitive functions 
of $M_\nu$, while $\eta_B$ is not in the RRL scenario considered
here,
strong correlations between the rates for these processes and $\eta_B$, 
found in new physics scenarios with other types of leptogenesis can be 
avoided.
\item
Except for this important message, several of the observations made by us 
with regard to the dependence of charged LFV processes on the complex phases
in the matrix $R$ and the Majorana phases
have been
already made by other authors in the rich literature on LFV
and leptogenesis. But most of these analyses were done
in the context of  supersymmetry. Here we would like to
emphasize that various effects and several patterns identified there
are valid also beyond low energy supersymmetry, even if
supersymmetry allows a definite realization of MLFV provided
right-handed neutrinos are degenerate in mass at the GUT
scale.
\item
 One of the important consequences of the messages above is the 
realization that
 the  relations between the flavour violating processes
       in the charged lepton sector, the low energy parameters
       in the neutrino sector, the LHC physics and the size of $\eta_B$ 
       are much richer in a general MLFV framework 
       than suggested by \cite{MLFV,CIP06}. Without a specific MLFV model 
       no general clear cut conclusions about the scale $\Lambda_{\rm LFV}$ on the 
       basis of a future observation or non-observation of $\mu\to e\gamma$
       with the rate $\ord(10^{-13})$ can be made in this framework.
\item
 On the other hand we fully agree with the point made in \cite{MLFV} that 
 the observation  of $\mu\to e\gamma$ 
        with the rate at the level
       of $10^{-13}$, is much easier to obtain within the MLFV scenario
          if the scales 
         $\Lambda_{\rm LFV}$ and 
         $M_\nu$ are sufficiently
       separated from each other. We want only to add that the necessary 
        size of this separation is sensitive to the physics
       between $M_Z$ and $\Lambda_{\rm GUT}$, Majorana phases and CP 
violation at high energy.
        In this manner
       the lepton flavour violating processes, even in the MLFV
       framework, probe scales well above the scales attainable
       at LHC, which is not necessarily the case within MFV in the quark
       sector.
\item
Finally, but very importantly, MLFV being very sensitive to new physics
at high energy scales, does not generally solve possible CP and flavour
problems. This should be contrasted with the MFV in the quark sector, 
where the sensitivity to new physics at scales larger than $1\tev$ is 
suppressed by the GIM mechanism.
\end{itemize}

\vspace{1cm} 

\noindent

{\bf Note}\\

\noindent
During the preparation of this revised version, one of us (S.U.) has
investigated parametric dependences in the present scenario
for the case of a real $R$ in more detail~\cite{Uhlig:2006xf}.

\noindent
{\bf Acknowledgements}\\
\noindent

We would like to thank P. Di Bari, M. Beneke, R.G. Felipe,
F.R. Joaquim, M. Pl\"umacher,
R. R\"uckl, M.A. Schmidt, F. Schwab, T. Underwood and E. Wyszomirski
for useful discussions.
We are grateful to S. Antusch and K. Turzynski for
  helpful comments.
This work has been supported by
Bundesministerium f\"ur
Bildung und Forschung under the contracts 05HT4WOA/3 and 05HT6WOA, the 
GIF project G-698-22.7/2002, the Humboldt foundation (G.C.B), and the
DFG Sonderforschungsbereich/Transregio 9 ``Computergest\"utzte
Theoretische Teilchenphysik'' (S.J., in part). The work of GCB is also supported by Fundacao para a Ciencia e a Tecnologia
( FCT Portugal ) through CFTP-FCT UNIT 777 and POCTI/FNU/44409/2002, 
POCI/FP/63415/2005. GCB would like to thank Andrzej Buras and his group for kind
hospitality at TUM.

\begin{appendix}
\section{Iterative solution of the renormalization group equations}
\label{app:iterative}
The goal of our numerical analysis of Section 4 is to determine the
neutrino Yukawa matrix $Y_\nu$ and the masses of the right-handed
neutrinos at the scale $M_\nu$ taking into account the
constraints on the masses and mixings of light neutrinos
measured at low energies and imposing the GUT condition
characteristic for the MLFV
\be
M_1(\Lambda_{\rm GUT})=
M_2(\Lambda_{\rm GUT})=
M_3(\Lambda_{\rm GUT}).
\ee
As discussed in Section~2 the latter condition implies
\be\label{conR}
{\rm Re}(R(\Lambda_{\rm GUT}))=0,
\ee
but $\text{Im}(R(\Lambda_{\rm GUT}))$ must be kept non-zero in order to have 
CP-violation
at high energy. 

The RG evolution from $\Lambda_{\rm GUT}$ down to $M_\nu$
generates small splittings between $M_i(M_\nu)$ and a non-vanishing
${\rm Re}(R(M_\nu))$, both required for the leptogenesis. 
As the splittings
between $M_i(M_\nu)$ turn out to be small, we integrate the right-
handed neutrinos simultaneously at $\mu=M_\nu$ imposing, up to their
splittings,
\be\label{OUT}
M_1(M_\nu)\approx
M_2(M_\nu)\approx
M_3(M_\nu)\approx M_\nu.
\ee
In view of various correlations and mixing under RG between
different variables we reach the goal outlined above by
means of the following recursive procedure:

{\bf Step 1}

We associate the values for the solar and atmospheric
neutrino oscillation parameters given in (\ref{In1})--(\ref{In2}) with the
scale $\mu=M_Z$ and set $\theta_{13}$ and the smallest neutrino mass 
$m_\nu^{\rm lightest}$  to
particular values corresponding to $\mu=M_Z$. 

{\bf Step 2}

For a chosen value of $M_\nu$, the RG equations, specific to a given 
MLFV model,
 are used to find
the values of the parameters of Step 1 at $\mu=M_\nu$. For instance
we find $m_i^\nu(M_\nu)$ and similarly for other parameters.

{\bf Step 3}

We choose a value for $\Lambda_{\rm GUT}$ and set first 
$m_i^\nu(\Lambda_{\rm GUT})=m_i^\nu(M_\nu)$ and similarly for
other parameters evaluated in Step 2. Setting next
\be\label{S2}
M_1(\Lambda_{\rm GUT})=
M_2(\Lambda_{\rm GUT})=
M_3(\Lambda_{\rm GUT})= M_\nu
\ee
and choosing the matrix R, that satisfies (\ref{conR}), allows also to 
construct 
$Y_\nu(\Lambda_{\rm GUT})$ by
means of the parametrization in (\ref{yukawaparamet}).

{\bf Step 4}

Having determined the initial conditions for all the
parameters at $\mu=\Lambda_{\rm GUT}$ we use the full set of the RG 
equations \cite{Antusch:2005gp} to
evaluate these parameters at $M_\nu$. In the range $M_\nu\le\mu\le \Lambda_{\rm
GUT}$  we use
\be
m_\nu(\mu)=-v^2\, Y^T_\nu(\mu) M^{-1}(\mu) Y_\nu(\mu).
\ee
The RG effects between $\Lambda_{\rm GUT}$  and $M_\nu$ will generally shift 
$m_i^\nu(M_\nu)$ to
new values
\be\label{S1}
\tilde m_i^\nu(M_\nu)=m_i^\nu(M_\nu)+\Delta m_i^\nu
\ee
with similar shifts in other low energy parameters. If these
shifts are very small our goal is achieved and the resulting
$Y_\nu(M_\nu)$ and $M_i(M_\nu)$ can be used for lepton flavour violating 
processes
and leptogenesis. If the shifts in question are significant
we go to Step 5.

{\bf Step 5}

The initial conditions at $\mu=\Lambda_{\rm GUT}$ are adjusted in order 
to obtain
the correct values for low energy parameters at $\mu=M_\nu$ as
obtained in Step 2. In particular we set
\be
m_i^\nu(\Lambda_{\rm GUT})=m_i^\nu(M_\nu)-\Delta m_i^\nu
\ee
with $\Delta m_i^\nu$ defined in (\ref{S1}).
Similar shifts are made for other parameters. If the condition
(\ref{OUT}) is not satisfied in Step 4, the corresponding shift in 
(\ref{S2})
should be made. Choosing $R$ as in Step 3 allows to
construct an improved $Y_\nu(\Lambda_{\rm GUT})$. 
Performing RG evolution with new
input from $\Lambda_{\rm GUT}$ to $M_\nu$ we find new values for 
the low energy
parameters at $M_\nu$ that should now be closer to the values
found in Step 2 than it was the case in Step 4. If necessary,
new iterations of this procedure can be performed until the
values of Step 2 are reached. The resulting $Y_\nu(M_\nu)$ and 
$M_i(M_\nu)$ are
the ones we were looking for.

\boldmath
\section{Basic Formulae for $\Delta_{ij}$}
\unboldmath
\subsection{Preliminaries}
In what follows we will present two generalizations of the formulae 
for $\Delta_{ij}$ in \cite{MLFV} in the approximation of degenerate 
right-handed neutrinos.  
We have checked that the splitting of $M_i^\nu$ by RGE has very small 
impact on these formulae. 
All the expressions for $\Delta_{ij}$ are meant to be valid at $M_\nu$.
Similar formulae have been given for instance in 
\cite{Petcov:2005yh,Kanemura:2005cq,Kanemura:2005it,Pascoli:2003rq,Petcov:2005jh},
 but we think that 
the formulae given below are more transparent.

In order to obtain transparent expressions for $\Delta_{ij}$ it is useful 
to introduce the mass differences
\be\label{dij}
\delta_{21}=m_{\nu 2}-m_{\nu 1}, \qquad \delta_{31}=m_{\nu 3}-m_{\nu 1},
\ee
\be\label{tdij}
 \tilde\delta_{21}(y_{12})=  \delta_{21}\cosh(2 y_{12}),\qquad
\tilde\delta_{31}(y_{12}) = \delta_{31}+m_{\nu 1}(1-\cosh(2 y_{12}))
\ee 
and collect the dependence on Majorana phases in the following two
functions
\be\label{F1}
F_1(\alpha,\beta)=e^{-i(\alpha-\beta)}+2ic^2\sin(\alpha-\beta),
\ee
\be\label{F2}
F_2(\alpha,\beta,\delta)=
s_{13} c^2 \cos
    (\alpha-\beta+\delta)+i \,c\,s  \sin(\alpha-\beta) 
+s_{13} s^2\cos(\alpha-\beta-\delta).
\ee

\boldmath
\subsection{$R$ real and $U_\nu$ complex}
\unboldmath
In this case we find
\begin{align}
\Delta_{\mu e}&=\frac{M_\nu}{\sqrt{2} v^2} (s ~c~ \delta_{21} +
e^{-i\delta }  s_{13} \delta_{31}), \label{DelEMuGone}\\
\Delta_{\tau e}&=\frac{M_\nu}{\sqrt{2} v^2} (-s ~c~ \delta_{21}
+ e^{-i\delta } s_{13}\delta_{31} ), \label{DelTauEGone}\\
\Delta_{\tau \mu}&=
\frac{M_\nu}{2 v^2} (-c^2\delta_{21} +
\delta_{31}), \label{DelTauMuGone}
\end{align}
where we have neglected terms $\ord(s_{13})$, whenever it was justified.

For the CP conserving cases $\delta=0,\pi$, these formulae reduce to the
formulae~(24) of \cite{MLFV} which represent the case where both
matrices, $R$ and $U_\nu$ are real. We note that in the presence of a real
matrix $R$, the $\Delta_{ij}$ do not depend on the Majorana phases and
$\Delta_{\tau\mu}$ does not depend on $\delta$. 
\boldmath
\subsection{$R$ and $U_\nu$ complex}
\unboldmath
Allowing for one additional phase $y_{12}$ in $R$, we find the
generalization of (\ref{DelEMuGone})--(\ref{DelTauMuGone}) that includes
CP violation both at low and high energies represented by $\delta \not
= 0,\pi$ and $x_{12},\,y_{12} \not = 0$, respectively

\begin{align}
\Delta_{\mu e}&=\frac{M_\nu}{\sqrt{2} v^2} 
   \left(s ~c~ \tilde\delta_{21}(y_{12}) +
e^{-i\delta }  s_{13} \tilde\delta_{31}(y_{12})
    + i \sqrt{m_{\nu 1} m_{\nu 2}} \sinh (2 y_{12})
       F_1(\alpha,\beta)\right),\label{DelEMuG}\\
\Delta_{\tau e}&=\frac{M_\nu}{\sqrt{2} v^2} 
   \left(-s ~c~ \tilde\delta_{21}(y_{12}) +
e^{-i\delta }  s_{13} \tilde\delta_{31}(y_{12})
    -i \sqrt{m_{\nu 1} m_{\nu 2}} \sinh (2 y_{12})
       F_1(\alpha,\beta)\right), \label{DelTauEG}\\
\Delta_{\tau \mu}&=\frac{M_\nu}{2 v^2} 
   \left(-c^2~ \tilde\delta_{21}(y_{12}) +
\tilde\delta_{31}(y_{12})
   +   2 i \sqrt{m_{\nu 1} m_{\nu 2}} \sinh (2 y_{12})
       F_2(\alpha,\beta,\delta)\right).
\label{DelTauMuG}
\end{align}
For $y_{12}=0$ (\ref{DelEMuG})--(\ref{DelTauMuG}) reduce to 
(\ref{DelEMuGone})--(\ref{DelTauMuGone}).
We note that relative to (\ref{DelEMuGone})--(\ref{DelTauMuGone}) there is 
an additional dependence on the difference of Majorana phases 
$\alpha-\beta$, collected in the functions $F_1$ and $F_2$ that disappears 
for $y_{12}=0$. This means that for $y_{12}$ very close to zero Majorana
phases in $l_i\to l_j\gamma$ decays do not matter but can be important 
already for small $y_{12}$.

Indeed, the $\Delta_{ij}$'s are very sensitive to
$y_{12}$ and the values of $\Delta_{ij}$ can be enhanced by several
 orders of magnitude 
\cite{Pascoli:2003rq,CAIB,Ellis:2002eh,Kanemura:2005cq,Kanemura:2005it,
Petcov:2005yh,Petcov:2005jh,Deppisch:2005rv}
relative to the case of $R=1$, even for
 $y_{12} =\ord(1)$.
Indeed as seen in (\ref{DelEMuG})-(\ref{DelTauMuG}), the $\Delta_{ij}$
depend exponentially on the $y_{12}$ and moreover for $y_{12} \not
= 0$, they do not only depend on the neutrino mass differences but
also on $\sqrt{m_{\nu 1}m_{\nu 2}}$ which can be much larger than
$\Delta m_{ij}$. Thus including a non-vanishing phase in $R$ can
have in principle a very strong impact on the analysis of \cite{MLFV} as also discussed in~\cite{CIP06}.

The large enhancement of $\Delta_{ij}$ in the case of a complex $R$ is
analogous to the large enhancement of $B(B_{d,s}\to\mu^+\mu^-)$ for large 
$\tan\beta$. In the latter case the presence of new scalar operators lifts 
the helicity suppression of the branching ratios in question. In the case of 
$\Delta_{ij}$ the appearance of a new mass dependence $m_i m_j$ in addition 
to $m_i-m_j$ has a similar effect provided $\sqrt{m_i m_j}\gg m_i-m_j$.
\end{appendix}

\renewcommand{\baselinestretch}{0.95}

\end{document}